\documentclass[twocolumn, preprint2]{aastex62}

\usepackage{float}
\usepackage{amsmath}
\usepackage{amssymb}
\usepackage[caption=false]{subfig}

\shorttitle{The case for a large-scale occultation network}
\shortauthors{Rice \& Laughlin}

\begin{document}


\title{The Case for a Large-Scale Occultation Network}


\author[0000-0002-7670-670X]{Malena Rice}
\affiliation{Department of Astronomy, Yale University, New Haven, CT 06511, USA}
\affiliation{NSF Graduate Research Fellow}

\author{Gregory Laughlin}
\affiliation{Department of Astronomy, Yale University, New Haven, CT 06511, USA}

\correspondingauthor{Malena Rice}
\email{malena.rice@yale.edu}



\begin{abstract}
We discuss the feasibility of and present initial designs and approximate cost estimates for a large ($N\sim2000$) network of small photometric telescopes that is purpose-built to monitor $V\lesssim15$ Gaia Mission program stars for occultations by minor solar system bodies. The implementation of this network would permit measurement of the solar system's tidal gravity field to high precision, thereby revealing the existence of distant trans-Neptunian objects such as the proposed ``Planet Nine." As a detailed example of the network capabilities, we investigate how occultations by Jovian Trojans can be monitored to track the accumulation of gravitational perturbations, thereby constraining the presence of undetected massive solar system bodies. We also show that the tidal influence of Planet Nine can be discerned from that of smaller, nearer objects in the Kuiper belt. Moreover, ephemerides for all small solar system bodies observed in occultation could be significantly improved using this network, thereby improving spacecraft navigation and refining Solar System modeling. Finally, occultation monitoring would generate direct measurements of size distributions for asteroid populations, permitting a better understanding of their origins.
\end{abstract}





\section{Introduction}
Stellar occultations are a long-standing observational technique used in a wide variety of applications. The rings of Uranus were serendipitously discovered during a stellar occultation of the planet that revealed five pre- and post-occultation dips in light produced by five of Uranus's narrow rings \citep{elliot1977}. Soon thereafter, the presence of Pluto's atmosphere was inferred from the detection of a symmetric, gently sloping lightcurve of the dwarf planet during a stellar occultation \citep{hubbard1988occultation, elliot1989}. The demonstration that Pluto's surface is not a mere airless expanse significantly augmented the motivation for an eventual mission. More recently, with the advent of \textit{Kepler} \citep{borucki2010} and a multitude of ground-based exoplanet detection surveys (e.g. HAT \citep{bakos2004}, WASP \citep{pollacco2006}, NGTS \citep{wheatley2017}), occultation-based transit photometry has constituted the most successful exoplanet detection method yet; data from exoplanet transit events has been used to measure planetary radii, atmospheric compositions, and system multiplicity \citep[e.g.][]{winn2015}. Further applications of occultation measurements range from ring seismology to study normal-mode planetary oscillations \citep[e.g.][]{baillie2011waves, 
mankovich2018cassini} to developing spatial maps of small solar system bodies \citep[e.g.][]{elliot2010size, sicardy2011pluto, alvarez2014stellar, buie2015size, leiva2017size}.


An intriguing use of occultations was recently demonstrated in support of the New Horizons Kuiper Belt Extended Mission, which included a flyby of Kuiper Belt Object (KBO) 486958 2014 MU69 (hereafter referred to as MU69) on January 1st, 2019 \citep{stern2018, Moore2018}. In order to coordinate this close-range flyby, it was necessary for New Horizons' navigation team to constrain the orbit of MU69 to exquisite precision. 

To accurately delineate the orbit, the team first set up a ``picket line" of twenty-five portable telescopes perpendicular to MU69's path to observe a predicted occultation on June 2nd-3rd, 2017 \citep{porter2018}. This initial campaign was unsuccessful; astrometric data of MU69 obtained with the Hubble Space Telescope (HST) in June and July indicated that the telescopes had been displaced by $\sim$80 km relative to the occultation path. Using this improved astrometry to update the predicted track, the team successfully observed an occultation  on July 10th, 2017 using the Stratospheric Observatory For Infrared Astronomy \citep[SOFIA;][]{young2012early} aircraft \citep{porter2018}. Shortly thereafter, on July 17th, 2017, the team employed another ``picket line'' of twenty-four 16-inch telescopes to capture an additional occultation, leading to further refinement of MU69's orbital elements and a direct measure of its size. Of the 24 telescopes used for the July 17th event, five successfully observed occultation chords across MU69's solid body \citep{porter2018}, while simultaneous HST photometry of the occulted star provided upper limits on additional material within the KBO's Hill sphere \citep{kammer2018probing}. A fourth occultation was subsequently predicted through orbital fitting in \citet{porter2018}, resulting in yet another successful observation on August 3rd, 2018.

The frequent occurrence of small telescope-observable occultation events (in this case, 3 events in 2 months) by an object of equivalent diameter $\sim$20 km that subtends only 600 $\mu$as, as well as the demonstrated success of precise astrometric measurements using arrays of small ground-based telescopes, motivates us to further investigate the technique's prospects. These observations of MU69 were by no means the first successful application of occultation measurements to study small bodies in the solar system; asteroid occultations have been captured with increasing frequency over time as stellar astrometry and asteroid ephemerides have grown more precise. The case of MU69 is, however, a particularly compelling proof-of-concept that, with a combination of precise stellar parallaxes from the Gaia Mission \citep{Gaia2016} and improved CCD and CMOS \citep{saint2017cmos} detector technologies, a relatively inexpensive network can be used to monitor occultations by distant minor planets with unprecedented precision.

In this paper, we explore the benefits of deploying a network of small, roboticized telescopes across the United States to monitor stellar occultations. Our proposed telescope grid is envisioned to use the infrastructure -- including site locations, access roads, communications provisioning, etc. -- associated with the ``cISP'' ultra-low latency microwave network that has been proposed for the continental United States \citep{Singla2015, Bhattacherjee2018}, thereby lowering costs and minimizing environmental impact. With a small telescope placed at each site on the cISP (see Figure \ref{fig:617patroclus}), the resulting network would span a vast area ($\sim$4000 km $\times$ 2500 km). In our view, the timing for such a project is right, with impetus amplified by the myriad of small solar system bodies that will soon be detected with the Large Synoptic Survey Telescope \citep[LSST;][]{Jones2016}.

We begin by providing a description of the network and its properties in \S\ref{section:network_overview}. Then, \S\ref{section:p9} outlines how the network can act as a gravitational probe for the putative ``Planet Nine'' \citep{batygin2016}. In the course of this overview, we explore the motivation behind a tidal acceleration search for Planet Nine (\S\ref{subsection:casetidalacc}), as well as the expected rate of observable occultation events (\S\ref{subsection:occultation_rates}), the refinement of orbital elements over time (\S\ref{subsection:oe_refinement}), and methods to distinguish the gravitational effect of Planet Nine from that of as-yet undetected bodies in the Kuiper belt (\S\ref{subsection:P9_v_KB}). \S\ref{section:applications} includes a discussion of various additional science cases for this network, including taking synergistic observations to complement upcoming NASA missions (\S\ref{subsection:NASA_synergies}); studying the morphology and size distribution of Jovian and Neptunian Trojan asteroids (\S\ref{subsection:trojans}); developing an acceleration map of the outer solar system using observations of Trans-Neptunian Objects (TNOs; \S\ref{subsection:TNOs}); and refining and/or obtaining asteroid diameter measurements (\S\ref{subsection:diameters}). \S\ref{section:camera_telescopes} involves an overview of our proposed instrumental setup based on comparisons with existing facilities, while \S\ref{section:multiplicity} addresses alternate potential versions of the network with fewer telescopes. We conclude with final remarks and a summary of our results in \S\ref{section:conclusions}.


\section{Network Overview}
\label{section:network_overview}
 The proposed cISP network, most recently described by \citet{Bhattacherjee2018}, consists of line-of-sight microwave relays connecting the 120 largest population centers in the continental United States. It is designed to provide an ultra-low-latency infrastructure for small packet traffic on the Internet, with individual links transmitting at data rates of order a gigabit per second. The cISP takes advantage of both optimized routing topology and the $\sim$40\% faster propagation speed of electromagnetic waves in air by comparison to glass fiber optic cable.  The cISP grid's design adopts only \textit{existing} communication towers in the FCC Antenna Structure Registration database,\footnote{\url{http://wireless.fcc.gov/antenna/index.htm?job=home}} and its sites are strategically placed at topographic high points chosen to confer optimal line-of-sight.  Construction of a cISP-based occultation-detection network would leverage access to these locations as well as the basic facilities -- including weather stations for relaying real-time meteorological conditions -- at each site.

\begin{figure*}
\centering
\includegraphics[width=0.95\textwidth]{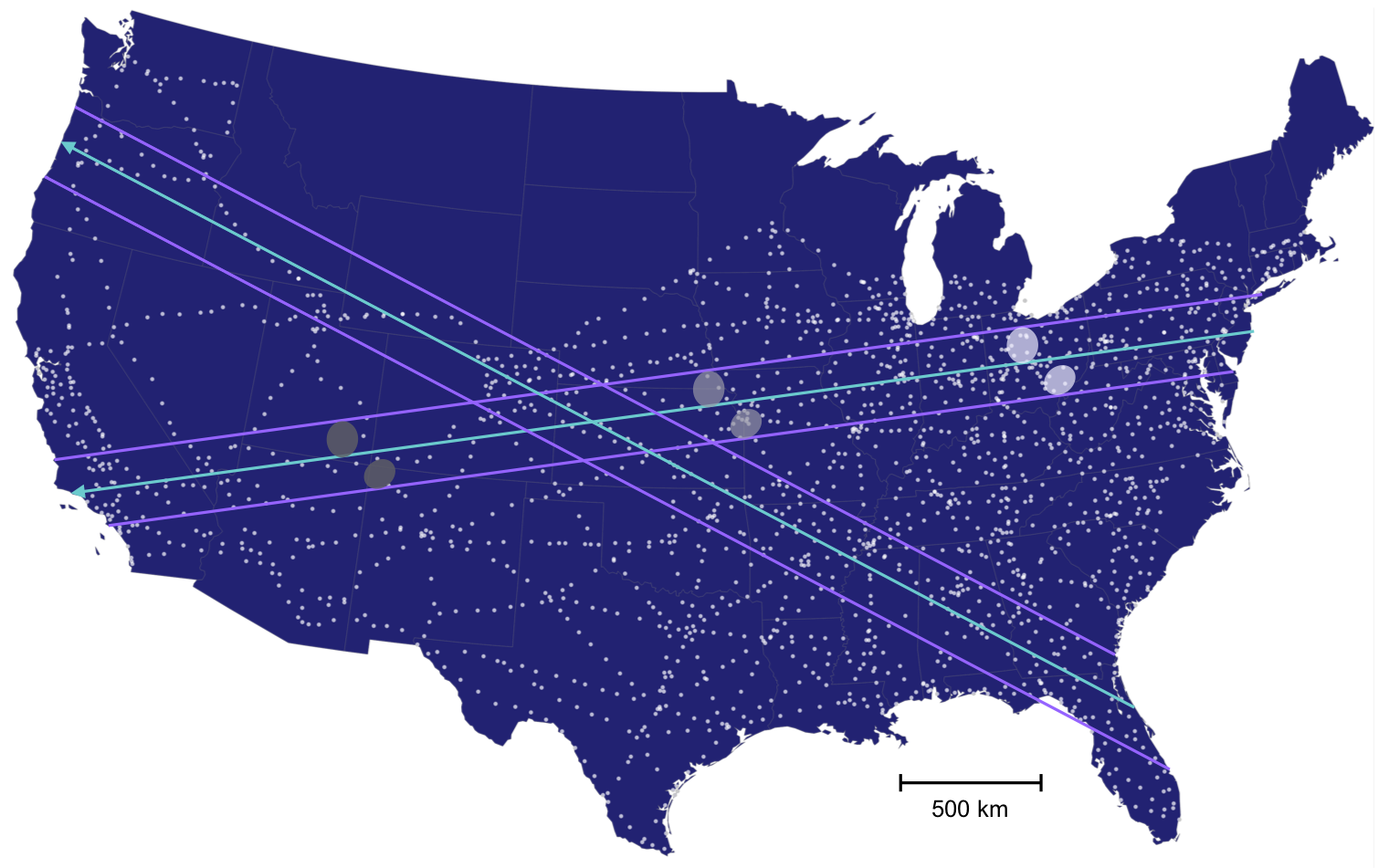}
\caption{Schematic of binary Jovian Trojan asteroid 617 Patroclus-Menoetius occulting the cISP network. Two possible occultation paths are shown to scale, bounded by purple lines with cyan arrows marking the central trajectories. Each white point on the map denotes a cISP network site, where the current cISP network design includes 1913 sites. State boundaries are shown in gray. Relative sizes of the two occulting bodies are based on results from \citet{buie2015size}, which reports axial sizes 127 x 117 x 98 km for Patroclus and 117 x 108 x 90 km for Menoetius from previous stellar occultation measurements.}
\label{fig:617patroclus}
\end{figure*}

In the system we envision, one roboticized telescope equipped with a high-cadence imager will be placed at each of $\sim$2000 points on the cISP network, resulting in an array that densely samples occultation tracks. For asteroids with diameter $\sim$10 km or larger, multiple telescopes will typically be activated as the shadow crosses the United States. An example is demonstrated in Figure \ref{fig:617patroclus}, which displays two sample trajectories of the binary Jovian Trojan 617 Patroclus-Menoetius -- a target of the \textit{Lucy} Jovian Trojan flyby mission \citep{levison2016lucy} -- overlaid on the current cISP network design, which includes 1913 sites. The uneven site spacing provides a balance between a large coverage area and higher-density regions necessary for detailed studies involving multiple observations of individual occultation events. Occultation tracks run predominantly from East to West from the perspective of ground-based observations near opposition; as a result, telescopes separated in the North-South direction by $D\gtrsim2\,{\rm km}$ will act as statistically independent accumulators of events. The Research and Education Collaborative Occultation Network \citep[RECON;][]{buie2016research} is an excellent example of a successful, currently-operating network of coordinated telescopes designed for occultation observations \citep[e.g.][]{benedetti2016results}, albeit on a much smaller scale than is proposed here and without roboticized operations.

\section{On the Detection of Planet Nine}
\label{section:p9}

\subsection{The Case for a Tidal Acceleration Search}
\label{subsection:casetidalacc}
Authors, notably including \citet{trujillo2014}, \citet{batygin2016}, and \citet{malhotra2016corralling}, have raised the possibility that the solar system contains an as-yet undetected planet orbiting far beyond Neptune (typical models often invoke an $M\sim10M_{\oplus}$ body on an eccentric orbit with $P\sim2\times10^{4}\,{\rm yr}$). Evidence for such a ``Planet Nine'' is indirect, however, and rests primarily on the larger-than-expected degree of orbital alignment among the most distant Kuiper belt objects. Additional clues are provided by the observed population of Centaur-like objects with high inclinations that are difficult to explain through alternative mechanisms \citep{batygin2016}.  Numerical simulations by \citet{holman2016observationalc}, \citet{millholland2017}, \citet{becker2017}, and \citet{batygin2017} (among others) have steadily refined the dynamical evidence and now point toward a planet with $m\sim5-10M_{\oplus}$ with semimajor axis $a\sim400-800\,{\rm AU}$. Parameter ranges based on current constraints are listed in Table \ref{tab:P9} \citep{batygin2019planet}, along with the assigned values used throughout this work.

Direct optical searches for Planet Nine in reflected sunlight are rendered difficult by the $r^{-4}$ decrease in brightness with distance, and they become extremely challenging for $V\gtrsim23-24$, especially in regions that are crowded with stars. Directed searches at infrared wavelengths \citep[e.g.][]{meisner2018} for self-luminous trans-Neptunian planets are stymied by the detection limits of wide-field space-based surveys (e.g. WISE \citep{wright2010} and NEOWISE \citep{mainzer2011}) unless a planet exists that is much larger and much warmer than expected. Because Planet Nine's emission is expected to peak at submillimeter wavelengths (50-100 $\mu$m), cosmic microwave background (CMB) surveys provide an alternative method to search for Planet Nine through its parallactic motion \citep{cowan2016cosmologists}. However, only a few current and upcoming CMB surveys have the requisite sensitivity \citep{cowan2016cosmologists}, and the effectiveness of this search method relies upon assumptions about Planet Nine's flux while also requiring a specific observing mode which repeatedly maps the same large regions of the sky at a cadence of a few months. Synergies between multiple detection methods may exist \citep{baxter2018planet}, but a definitive assessment remains elusive. Furthermore, in the case that a ninth planet \textit{does not} exist, it is difficult to rule out its presence simply through the lack of a direct detection.

Fortunately, however, indirect detection via the planet's tidal gravity signature may be feasible. Shortly after \citet{batygin2016} issued their detailed prediction of the Planet Nine orbit, \citet{fienga2016} combined a high-resolution solar system dynamical model with telemetry from the Cassini Mission to assess whether the data and model are consistent with the presence of Planet Nine at different locations along \citet{batygin2016}'s fiducial orbit. They found no constraint on the planet near apoastron while simultaneously determining that its presence near periastron is inconsistent with the data. This work has since been revisited by \citet{folkner2016}, who found that, under certain assumptions regarding the mass distribution in the Kuiper belt, perturbing bodies larger than 10$M_{\oplus}$ at $1000\,{\rm AU}$ are disallowed. \citet{holman2016observationalp, holman2016observationalc} further analyze the proposed Planet Nine's effects on the known solar system ephemeris, concluding that models are best fit where Planet Nine is placed at or near apoastron in its orbit.

\begin{table}[]
    \centering
    \caption{\small{\textbf{Planet Nine}}}
    \begin{tabular}{c|c|c}
        \hline
        Parameter & Allowed Range & Assigned Value \\
        \hline
         a & 400-800 AU & 600 AU \\
         e & 0.2-0.5 & 0.35 \\
         i & 15-25$\degr$ & 20$\degr$ \\
         m & 5-10M$_{\oplus}$ & 7.5M$_{\oplus}$ \\
        \hline
    \end{tabular}
    \vspace{2mm}
    \label{tab:P9}
\end{table}

The instantaneous tidal differential acceleration $a_{t}$ induced by a perturbing object of mass $m$ at a distance $d$ from the Sun is given by

\begin{equation}
    a_{t}=\frac{x_B G m}{d^3}\, ,
\end{equation}

\hspace{-4mm} where $x_B$ is the effective baseline across which the tidal acceleration is applied. For an as-yet undetected 7.5$\, M_{\oplus}$ object at $d=810\,{\rm AU}$ (the apoastron Sun-Planet Nine distance corresponding to the Planet Nine properties adopted from Table \ref{tab:P9}) and measured with a baseline $x_B\sim10\,{\rm AU}$ appropriate to Jupiter's Trojan asteroids,  $a_t \sim 3\times10^{-13}\,{\rm cm\,s^{-2}}$, corresponding to a displacement $dx \sim \frac{1}{2} a_t t^2 \sim 30\,{\rm m}$ over a 5-year time scale.

A net displacement of order $dx\sim 30 \,{\rm m}$ incurred over 5 years may, at first glance, appear to be insignificant, but it has the same order of precision to which solar system ephemerides are currently determined.  For example, the range residuals to the Cassini orbiter's telemetry reported by \citet{fienga2016} are of order $dx \sim 10-100$ m, whereas the Martian range has been determined to within 3.6 m using just over one year of Mars Express data, and the lunar range to within 2 cm using 10 years of Apollo retro-reflector data from the Apache Point Observatory and 2 years of infrared lunar laser ranging data from the Grasse station \citep{viswanathan2017, viswanathan2018lunar}.

Occultation monitoring can be used to measure the accumulated effects of the tidal acceleration imparted by massive outer solar system objects, including Planet Nine. To fix ideas, we investigate the use of the aggregate Jovian Trojan populations, located at the L4 and L5 Jupiter-Sun Lagrange points, as a probe of tidal gravity. This population provides an abundance of available objects to track -- a significant statistical sample that permits triangulation to distinguish perturbations of the Earth's orbit from those of the Trojans' orbit -- as well as a relatively long baseline over which the differential tidal acceleration from an external perturber can be measured ($\sim$10 AU across a Jovian Trojan orbit). Over 7000 Jovian Trojans are currently catalogued in the JPL Small-Body Database,\footnote{\url{https://ssd.jpl.nasa.gov/}} and this number is anticipated to surge with the advent of LSST. Both Jovian Lagrange points are well-populated, with number asymmetry $N(\mathrm{L}_4)/N(\mathrm{L}_5) = 1.34$ reported from WISE/NEOWISE observations \citep{marzari2002origin, grav2012}. Hereafter in \S\ref{section:p9} we use the terms ``Trojan" and ``Jovian Trojan" interchangeably.

An accurate measurement of tidal acceleration must combine precise astrometry with a rigorous dynamical model of the full solar system to integrate the trajectories of occulting objects. Together, the combination of measurements and model enables the prediction of future sky positions that can be compared with future observations. The requisite stellar astrometric precision is provided by the Gaia Mission, while solar system ephemeris models \citep[e.g.][]{giorgini1996jpl, fienga2014inpop, folkner2014planetary, pitjeva2018} combined with occultation measurements provide ever-stricter constraints on the projected positions of foreground asteroids over time.

The aforementioned Gaia Mission is obtaining accurate sky positions and sky motion models for $N\sim10^{9}$ stars with $G<20.0$, where Gaia's G filter is similar to the standard V-band filter \citep{Gaia2016}. For this large aggregate of stars, the original mission specifications mandated that 50\% of sources have $G<19$ and that 2.27\% ($\sim$26 million stars) have $G<14.5$. The actual astrometric precision to which Gaia's observed stars can be fixed varies inversely with brightness, with design expectations ranging from $\omega=7\,\mu$as at $V=10$ to $\omega=12-25\, \mu$as at $V=15$ and $\omega=100-300 \,\mu$as at $V=20$, where the ranges at a given magnitude depend upon the intrinsic brightness of the star \citep{de2005gaia}. \citet{gaia2018} have recently published Gaia's Data Release 2 (DR2), an interim compendium of measurements based on 22 months of spacecraft operations.

Most small bodies in the solar system have ephemerides determined from photometric measurements which, in the absence of radar range measurements, have accuracy tied to the pixel scale, $p$, of the employed imaging detector.\footnote{\url{https://www.minorplanetcenter.net/}} For individual observations, this is typically of order $p=1\arcsec/{\rm px}$, corresponding to sky plane displacements of $\sim 3,700\,{\rm km}$ at the distance of the Trojan asteroids and $\sim 30,000\,{\rm km}$ for objects in the Kuiper belt. When a minor planet occults a star, however, its instantaneous ephemeris is radically improved. An observed blocking of starlight indicates that a chord traversing the line of sight to the object also traversed the star, thereby establishing a time accuracy $\Delta t \sim \delta/v_T$ where $\delta$ is the angle subtended by the transit chord and $v_T$ is the angular velocity of the body on the sky. The spatial accuracy on the plane of the sky is of order $\omega$, the astrometric accuracy of the star's position. With the advent of Gaia DR2, the improvement factor from previous Hipparcos astrometric uncertainties is often measured in the thousands \citep{perryman1997hipparcos, lindegren2018}. A Trojan asteroid observed to occult a bright Gaia star generates a one-time sky plane positional accuracy of just a few tens of meters, comparable to the order-of-magnitude tidal displacement $dx\sim30\,{\rm m}$ induced by a Planet Nine-like object over a half-orbit time scale for the Trojan occulter. This startling potential accuracy is the primary motivator for the investigation presented here.

\subsection{Occultation Rates}
\label{subsection:occultation_rates}

The population of Jovian Trojan asteroids is understood to be extensive, with estimates ranging from $N\sim1\times10^4$ to $N\sim2\times10^{5}$ objects with $D>2\,{\rm km}$ in the combined $L_4$ and $L_5$ swarms \citep{jewitt2000, yoshida2005size, fernandez2009, wong2015color}. Variations in $N$ stem from assumptions regarding the underlying albedo distribution. The Gaia DR2 data release, moreover, contains of order $3\times10^{7}$ stars of apparent brightness $V<15$ whose sky positions have been determined to accuracy $\omega \sim 20 \mu{\rm as}$ \citep{de2005gaia}. At typical Trojan distance $a=5.2 \,{\rm AU}$, this is equivalent to sky-plane positional accuracy $dx\sim 75\,{\rm m}$ for the portion of the asteroidal body responsible for the occultation. 

To estimate the rate of observable Trojan occultations, we must first assume a Trojan size distribution. We adopt the broken power law differential magnitude distribution 

\begin{equation}
    \frac{dN}{dH} = 
        \begin{cases}
        10^{\alpha_1(H-H_0)} & H < H_b \\
        10^{\alpha_2H + (\alpha_1 - \alpha_2)H_b - \alpha_1H_0} & H \geq H_b
        \end{cases}
\label{eq:dN_dH}
\end{equation}

\begin{figure}
    \centering
    \includegraphics[width=0.45\textwidth]{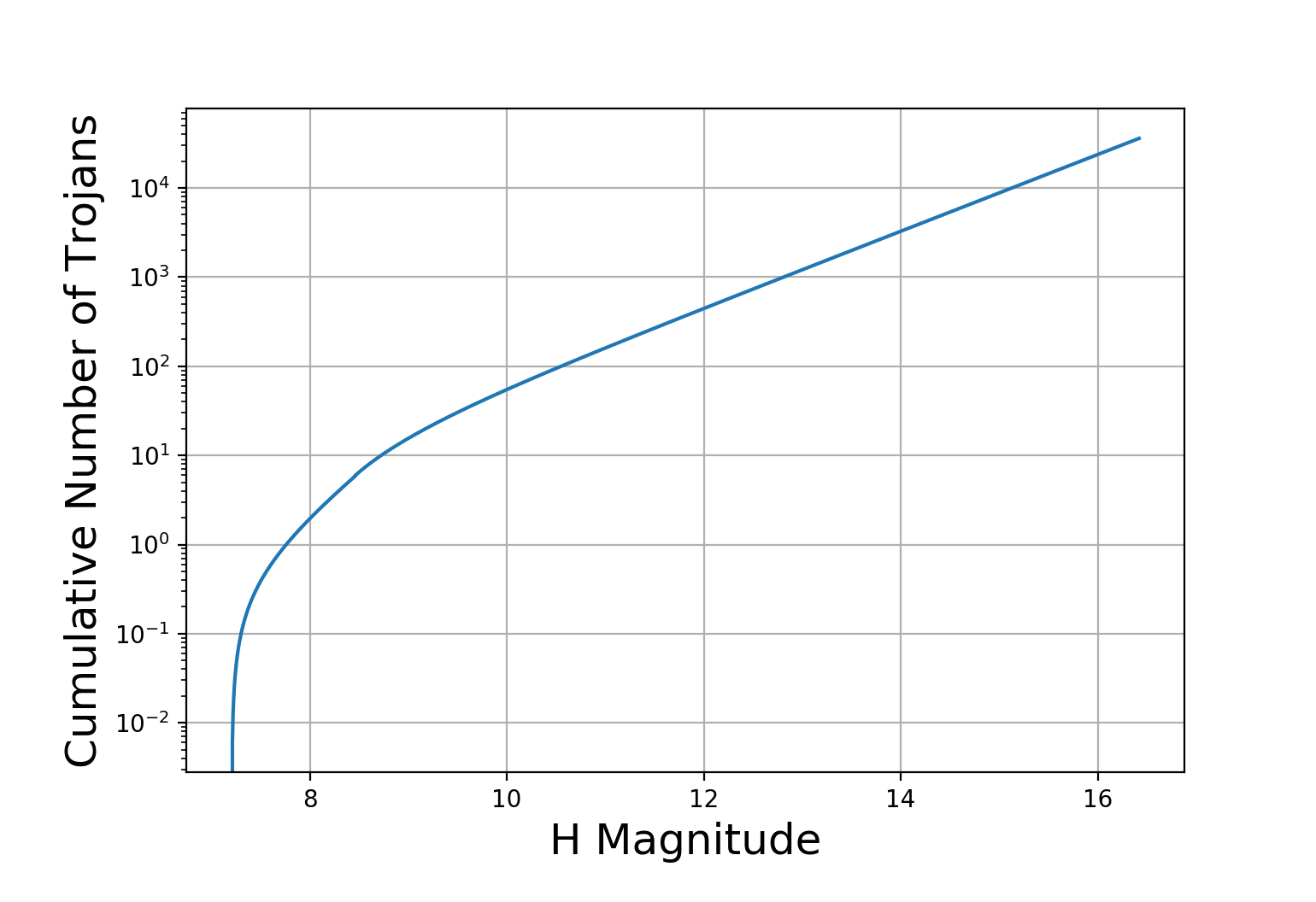}
    \caption{Cumulative number of Trojans as a function of $H$ magnitude, estimated from Equation \ref{eq:dN_dH} using parameters obtained in \citet{wong2015color}.}
    \label{fig:Hmag_v_ntroj}
\end{figure}

\hspace{-4mm} and associated best-fitting parameters $\alpha_1 = 0.91$, $\alpha_2=0.43$, $H_0=7.22$, and $H_b=8.46$, obtained in \citet{wong2015color} by fitting the distribution of bright Trojans. The resultant magnitude distribution is shown in Figure \ref{fig:Hmag_v_ntroj}. We combine this magnitude distribution with Equation \ref{eq:H_to_rad} to obtain the corresponding size distribution, where we use the process described in Appendix \ref{section:troj_diameters} to set albedo values.

We then sum the diameters of all inferred Trojans with $D>2$ km ($7.2 < H < 16.42$) to obtain a total length $1.4\times10^5$ km for all $3.6\times10^4$ large Trojans lined up side-by-side. Given $P\sim12\,{\rm yr}$ orbital periods, each Trojan sweeps out $\sim$30 degrees yr$^{-1}$, and the aggregate of Trojans sweeps out projected area $\sim$0.31 square degrees yr$^{-1}$. For a uniform distribution of $V<15$ stars on the sky, this corresponds to $N\sim225$ potentially observable occultations per year at each location on Earth. Using a network of $N\sim2000$ small telescopes, approximately $4.5\times10^{5}$ events are thus observable annually, some of which may not be unique -- i.e. multiple telescopes may observe a single occultation event. Assuming an $f=0.6$ duty cycle to account for airmass, solar angle, and overlapping targets, as well as an additional $f=0.4$ duty cycle when considering the day-night cycle, one plausibly expects $\sim1.1\times10^5$ unique occultation events per year on the network, of which a subset would be observable by multiple telescopes.

Focusing on the catalog of known Trojans in the JPL Small-Body Database, we provide a more detailed assessment to compare with the foregoing estimates. The occultation rate for an individual Trojan is 

\begin{equation}
    r = n_* \sigma v_{T},
\end{equation}

\hspace{-5mm} where $n_*$ is the areal number density of background stars, $\sigma$ is the linear cross-section for occultations, and $v_{T}$ is the angular speed of the Trojan relative to the observer at the time of occultation. For the purpose of assessing rates, rather than planning specific observations, we adopt a simplified solar system model containing only the Earth, the Sun, and the Trojan of interest. We ignore Earth's axial precession and assume that background stars are point sources, which is an appropriate approximation at the scale of this problem. For example, a G2 solar-type dwarf star located at $d=1000\,{\rm pc}$ has $V\sim15$ and angular size $\delta\sim9.3\,\mu$as, smaller than the projected end-of-mission $\omega=12-25\, \mu$as astrometric precision of Gaia.

To study the frequency of observable occultation events, we simulate the Sun-Earth-Trojan system for each of the 7207 Jovian Trojans listed in the JPL Small-Body Database on November 1st, 2018. The starting spatial distribution of these Trojans relative to Jupiter and the Sun in our simulations is shown in Figure \ref{fig:trojans_distribution}. Combining the \texttt{REBOUND} orbital integrator \citep{rein2012rebound} with standard coordinate frame transformations, we obtain the RA/Dec coordinates for each Trojan at 2000 equally spaced time steps over 24 years -- approximately the duration of two full Trojan orbits. We calculate the expected occultation rate for each Trojan at each time step, with each component of the rate equation obtained using the following procedure:



\begin{figure}
    \centering
    \includegraphics[width=0.47\textwidth]{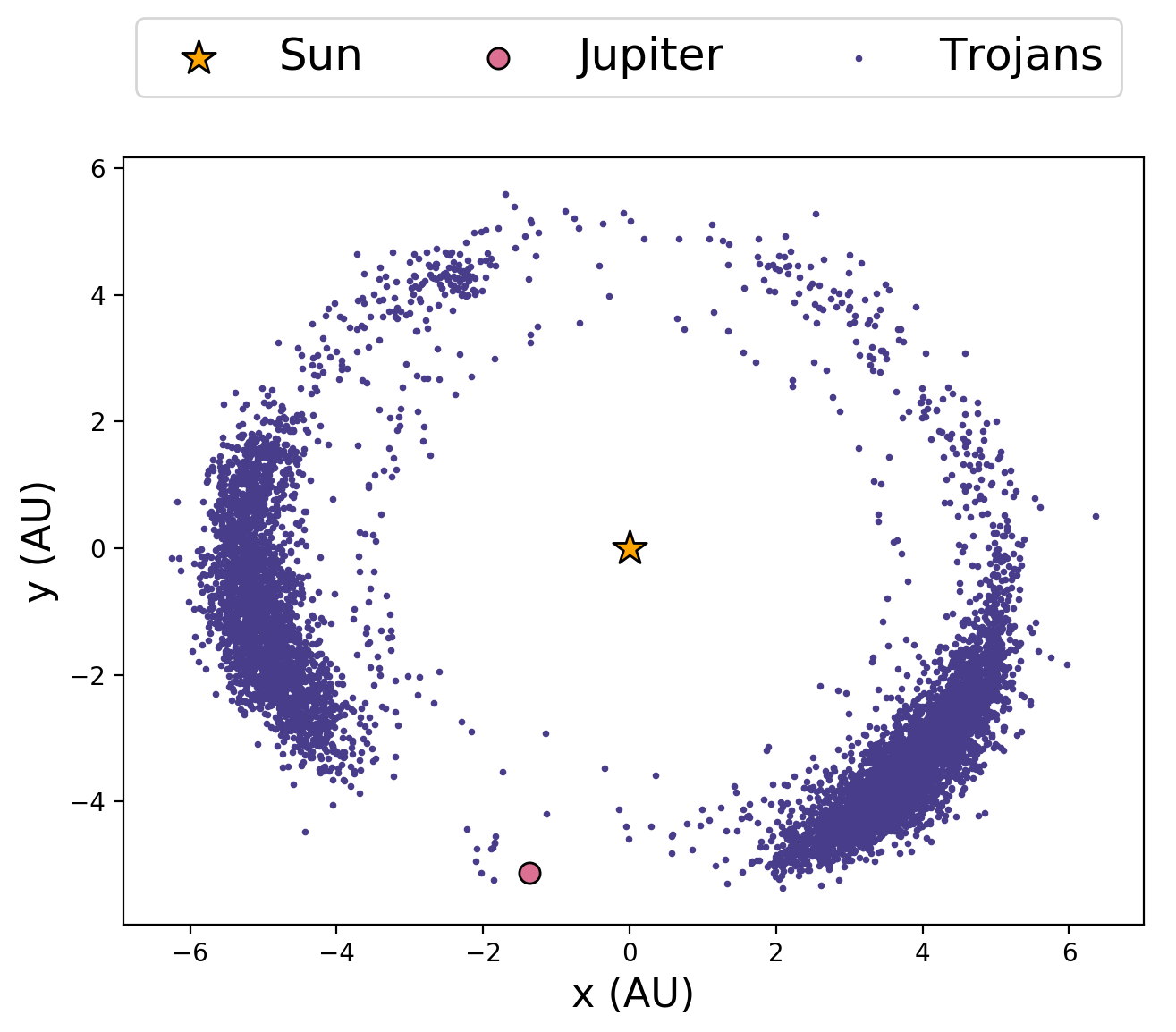}
    \caption{Distribution of all modeled Jovian Trojan asteroids at time t=0 in our occultation rate calculation.}
    \label{fig:trojans_distribution}
\end{figure}

\begin{figure*}
    \centering
    \includegraphics[width=1.0\textwidth]{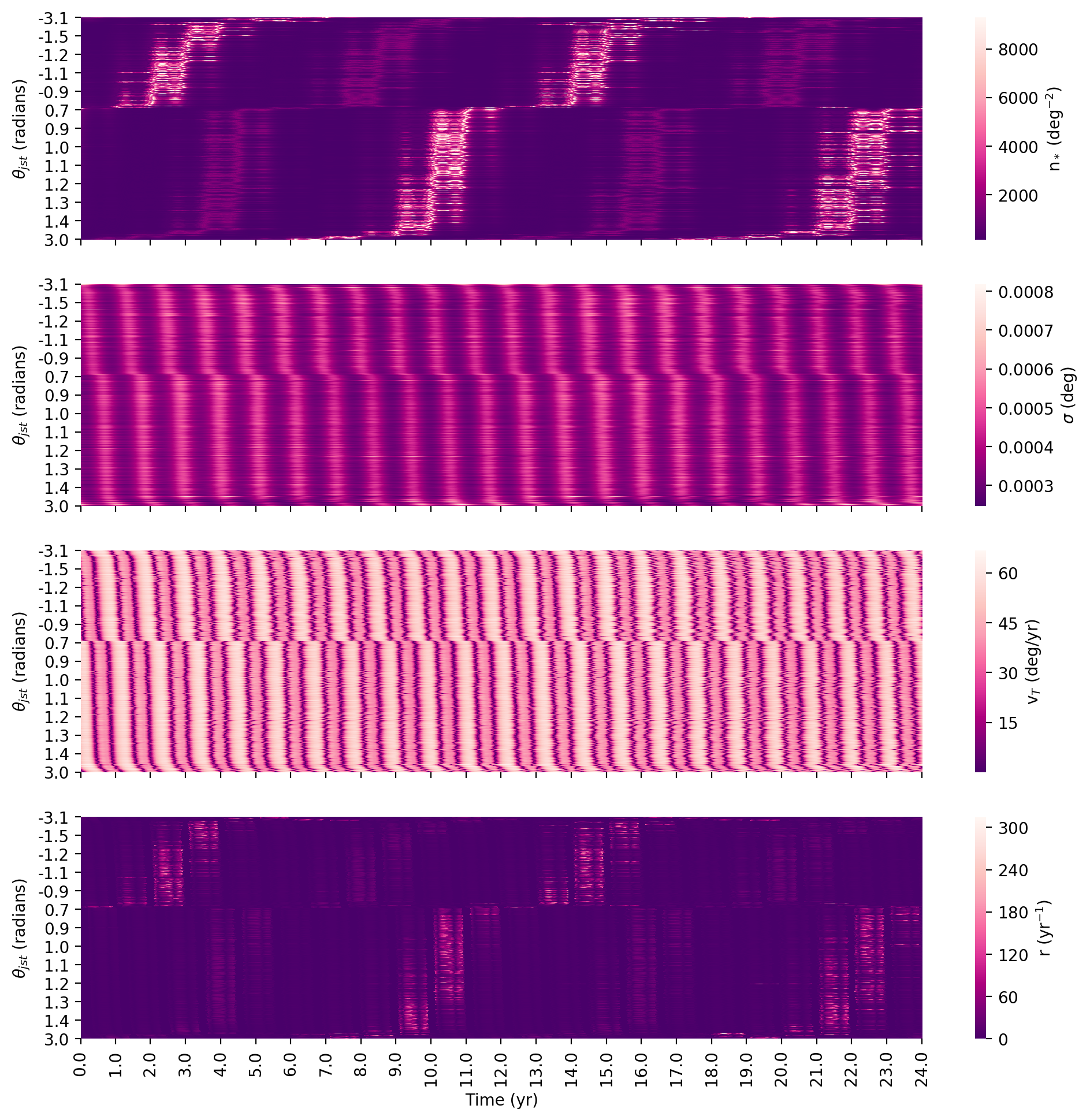}
    \caption{$n_*$, $\sigma$, $v_T$, and occultation rate $r$ displayed as a function of time for all 7207 Trojan asteroids in the JPL Small-Body Database. Trojans are ordered by angle $\theta_{jst}$ between the Sun-Jupiter and Sun-Trojan vectors. The distribution of occultation rates in the bottom map consists of the top 3 maps multiplied together, with an additional condition that regions with $\theta_{set} < 38\degr$ have occultation rate set to zero, where $\theta_{set}$ is the angle between the Earth-Sun and Earth-Trojan vectors.}
    \label{fig:numtroj_occrate}
\end{figure*}

\begin{figure}
    \centering
    \includegraphics[width=0.47\textwidth]{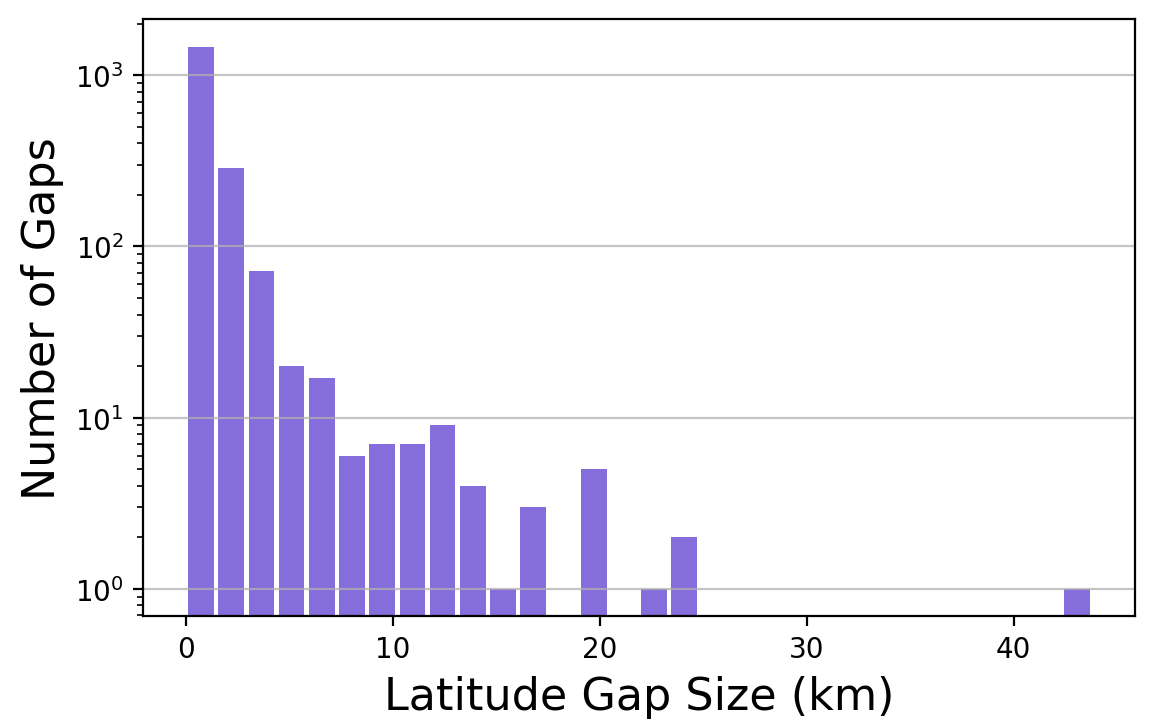}
    \caption{Distribution of all latitudinal gaps within the current cISP network setup. These gaps are obtained by ordering all network site locations by latitude and finding the gap size between all adjacent sites in this listing.}
    \label{fig:network_gaps}
\end{figure}

\begin{itemize}
    \item $n_*$: We categorize the total number of stars along the ecliptic plane (to rough approximation, the orbital plane of the Jovian Trojans) into 360 RA bins that are evenly spaced such that each bin encompasses 1$\degr$ in RA space. We choose to parameterize over RA space in order to encompass regions predominantly passing through the Galactic plane within individual bins.
    
    To accomplish this, we advance Jupiter's orbit at 360 evenly spaced time steps (corresponding to slightly varying step sizes in RA space) using \texttt{REBOUND}. Between each set of adjacent RA values, we query the Gaia DR2 database within $\pm2.5\degr$ from the midpoint declination value to obtain a star count within each bin. Using these total star counts and the size of each bin on the sky, we calculate the stellar density in each bin. From this point, we linearly interpolate our results to obtain stellar densities in evenly spaced RA bins. We obtain $n_*$ at each time step by selecting the RA bin within which the Trojan falls at that time, then retrieving the corresponding stellar density.
    
    \item $\sigma$: The effective occultation cross-section for each Trojan can be estimated as the 2545 km latitudinal extent of the continental United States subtended at the instantaneous Earth-Trojan distance. This cross-section takes into account the full range of telescopes extending across the country such that nearly all asteroids with $D\gtrsim 2$ km passing across the United States in the East-West direction can be seen in occultation by one or more network telescopes. 
    
    Among the latitudinal distribution of telescopes, gaps larger than the diameter of the occulter may prevent the detection of an occultation, since small asteroids can slip through these gaps. Few large latitudinal gaps (33 gaps larger than 10 km in extent) exist within the cISP network design that we employ, and the full distribution of gaps is displayed in Figure \ref{fig:network_gaps}. For the initial estimates presented here, we do not explore this in further detail but acknowledge that these gaps may lead to an underestimate of the total occultation rate. If desired, these gaps could be addressed by adding new, strategically placed telescope sites to the current network design, increasing the total number of sites from $N=1913$ to $N\sim2000$.

    \item $v_T$: This is the effective angular velocity of the Trojan, which is given by the absolute value of the sky-plane velocity of the Trojan relative to that of the Earth ($\mid$\textbf{v$_E$} - \textbf{v$_T$}$\mid$) at each time step.
    
\end{itemize}

Our resulting $n_*$, $\sigma$, and $v_T$ values, as well as the corresponding occultation rates $r$, are displayed in Figure \ref{fig:numtroj_occrate}. All four maps order the Trojans by $\theta_{jst}$, the angle between the Sun-Jupiter and Sun-Trojan vectors, where Figure \ref{fig:trojans_distribution} provides the orientation of the full system. We use the convention that negative angles correspond to Trojans lagging behind Jupiter in its orbit (the L5 swarm), while positive angles correspond to Trojans leading Jupiter (the L4 swarm). 

The structure in the top panel of Figure \ref{fig:numtroj_occrate} reflects the periodicity with which the two Trojan swarms pass through the Galactic midplane. Bright maxima in the background stellar density $n_*$ of each swarm occur with a $\sim$12-year periodicity corresponding to the frequency with which they pass through the Galactic midplane in the direction of the central Milky Way. Dimmer local maxima occur 6 years apart from each of these brighter maxima, corresponding to passage through the outer region of the Galactic plane with a lower stellar density. The offset between central peaks in L4 and L5 is approximately 4 years in accordance with the spatial separation of the two swarms along Jupiter's orbit.

The second panel shows the linear cross-section $\sigma$ of the United States as viewed by each Trojan asteroid as it moves through its orbit. Alternating bright and dark streaks, each with a duration of $\sim6$ months, reflect the Earth's periodic movement towards and away from the two Trojan swarms during its orbit.

The third panel displays the sky-plane velocity $v_T$ of each Trojan as a function of time, where dark streaks correspond to times during which the Trojan's motion is perpendicular to Earth's sky-plane (approximately twice per Earth orbit). Between these dark streaks are alternating high- and medium-velocity local maxima. High-velocity local maxima correspond to times during which the Earth and the Trojan move in opposite directions within their orbits (for example, when they are on opposite sides of the Sun), increasing the Trojan's relative velocity as viewed from Earth. Conversely, medium-velocity local maxima occur where the Earth and the Trojan move in the same direction, such as when they are both on the same side of the Sun, reducing the Trojan's relative sky-plane velocity.

Finally, our fourth panel is produced by multiplying together the first three panels and incorporating an additional condition that all regions with $\theta_{set} < 38\degr$ have occultation rate set to zero, where $\theta_{set}$ is the angle between the Earth-Sun and Earth-Trojan vectors. This final condition accounts for solar angle and airmass considerations, where we require that, for a Trojan to be observable during the night, (1) the solar angle must be -8$\degr$ or lower relative to the horizon, slightly below civil twilight \citep{nichols1964}, when the Trojan is in the sky, and (2) the asteroid must be at airmass $z<2$, or placement in the sky at least 30$\degr$ above the horizon. We assume that all Trojans are observable from the continental United States due to the orientation of their orbits roughly along the solar system ecliptic plane.

The overall structure of occultation rates in Figure \ref{fig:numtroj_occrate} follows that of $n_*$, with overlaid dark streaks reflecting the influence of $\sigma$ and $v_T$. Longer-lasting dark streaks are superimposed as well due to our requirement that $r=0$ where $\theta_{set} < 38\degr$.

\begin{figure}
    \centering
    \includegraphics[width=0.45\textwidth]{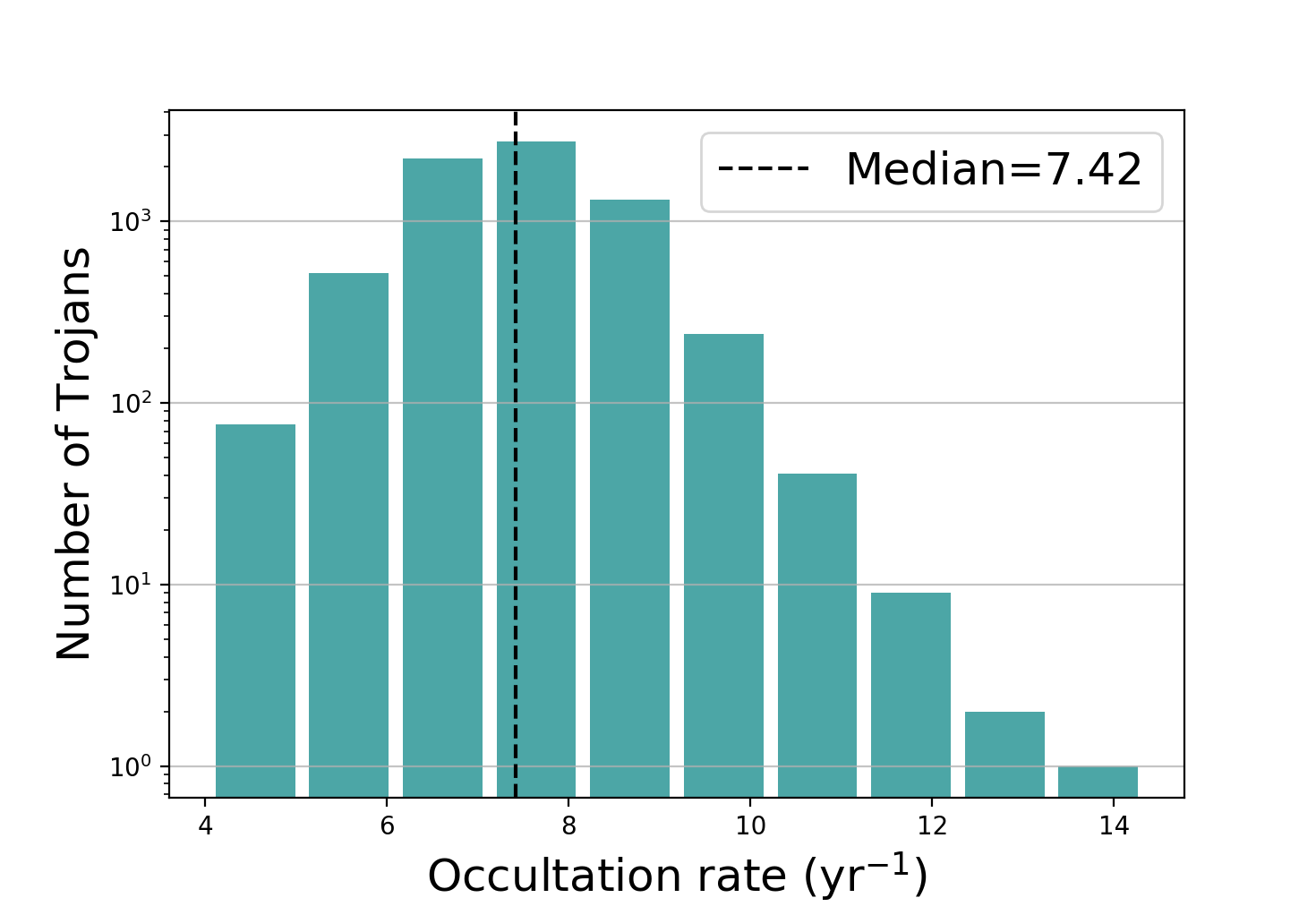}
    \caption{Average occultation rates obtained for each Trojan over 2000 sampled time steps spanning 2 years. The median number of total occultations per year is 7.42.}
    \label{fig:occ_rate_hist}
\end{figure}

From this distribution, we extract a median rate of 7.42 occultations per Trojan per year, with the rate distribution displayed in Figure \ref{fig:occ_rate_hist}. Extrapolating to the full Jovian Trojan population with $D>2$ km (N $\sim 3.6\times10^4$), this corresponds to $\sim2.7 \times 10^5$ total occultations per year. Some of these events will occur during the daytime and thus will not be observable. Approximating that, on average, 9/24 hours per day can be spent observing, we estimate a total of $1.0\times 10^5$ observable occultation events per year, in excellent agreement with our previous estimate. This is a sufficiently high occultation rate to regularly track a large number of Trojan asteroids in search for signatures of a massive perturber. We envision that, in conjunction with the proposed program to track Trojan asteroids, the network will be kept continually busy with observations of various small-body occultations throughout the solar system.

Although this estimate does not account for weather conditions, we argue that, for all but the smallest targets, the extended nature of this network will mitigate weather problems. An occulting body will cross numerous network sites as it traverses the continent, making it unlikely for weather conditions to prevent observations by $\textit{all}$ telescopes along the path. In particular, the characteristic Rossby parameter ($\beta=2\omega\cos \phi /a$, where $\phi$ is the latitude, $\omega$ is Earth's angular rotation speed, and $a$ is Earth's radius) at mid-latitudes generally ensures nonuniform weather conditions across the country. For the smallest targets, which occult only 1-2 sites, however, weather conditions will play a larger role.

\subsection{Orbital Element Refinement}
\label{subsection:oe_refinement}
A set of three occultation measurements over a time baseline of several months or more permits, in idealized principle, the determination of all six osculating Keplerian orbital elements to the precision of the measured sky-plane positions \citep{bernstein2000orbit}. To numerically evaluate the precision to which we can pin down the orbit of a Trojan with a given number of occultation measurements, we simulate a series of successive occultations and study their effect on refining the orbital elements of the occulter. We report results for medium-sized Jovian Trojan 3794 Sthenelos from the JPL Small-Body Database, with properties summarized in Table \ref{tab:3794sthenelos}.

\begin{table}
    \centering
    \caption{3794 Sthenelos}
    \begin{tabular}{c|c|c}
    \hline
        Parameter & Value & Uncertainty \\
    \hline
         a & 5.204 AU & 1.246e-7 AU \\
         e & 0.147 & 6.766e-8 \\
         i & 0.106$\degr$ & 1.163e-7$\degr$ \\
         $\Omega$ & 5.990$\degr$ & 9.742e-7$\degr$ \\
         $\omega$ & 6.215$\degr$ & 1.052e-6$\degr$ \\
         M$_0$ & 5.355$\degr$ & 5.122e-7$\degr$ \\
         D & 34.531 km & 0.364 km \\
         \hline
    \end{tabular}
    \vspace{2mm}
    \label{tab:3794sthenelos}
\end{table}

\begin{figure}
    \centering
    \includegraphics[width=0.47\textwidth]{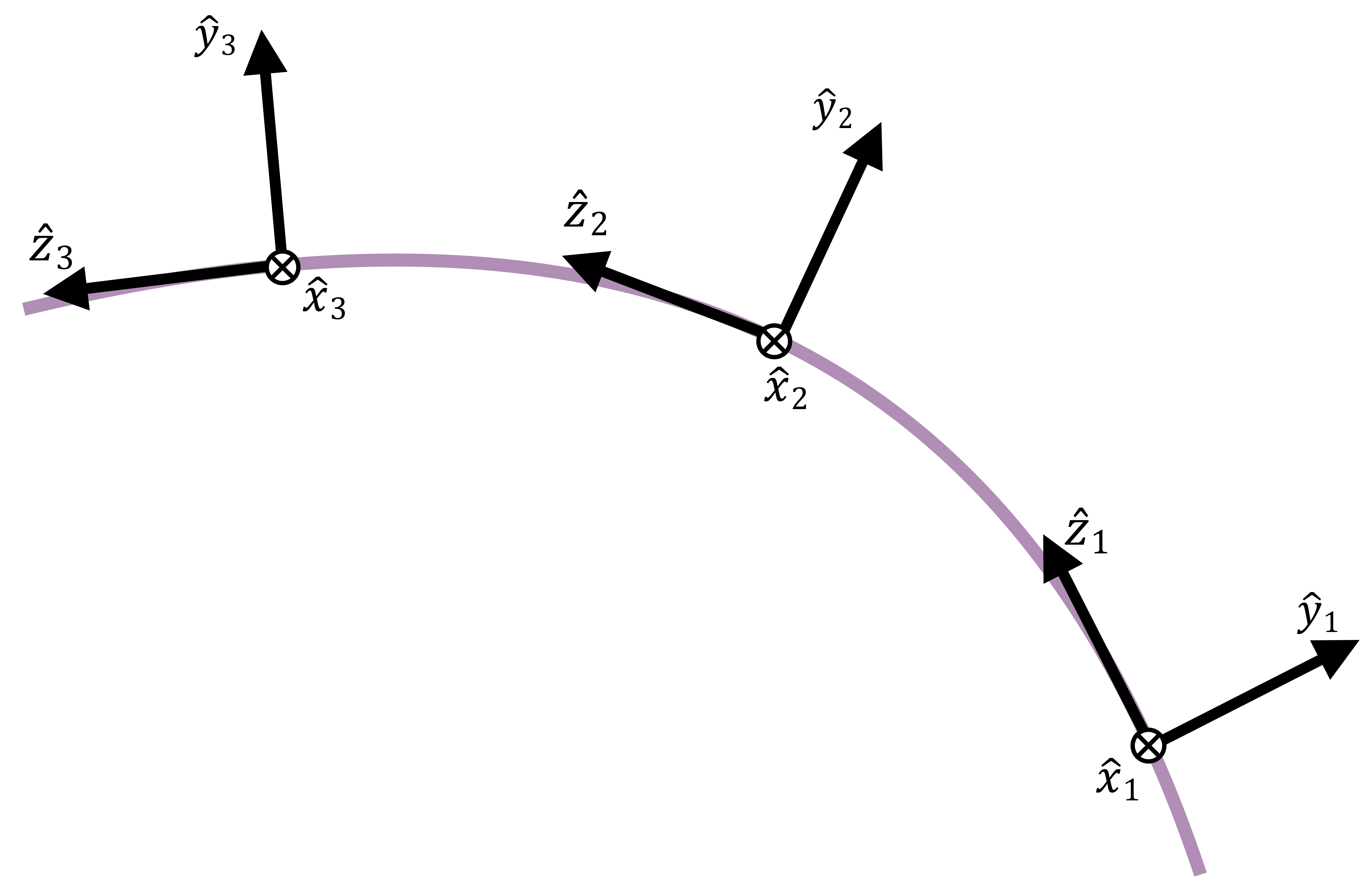}
    \caption{Schematic of coordinate frame transformations for three occultations. Each $\hat{x}$ is in the observer-Trojan direction, $\hat{z}$ is in the direction of the Trojan's sky-plane velocity, and $\hat{y}$ is in the mutually perpendicular direction.}
    \label{fig:coordframes}
\end{figure}

We first assume that the fiducial reported orbit from the JPL Small-Body Database is the `true' orbit. We select $N_{occ}$ points along the orbit at which we simulate occultations, and for each of these points we obtain a separate coordinate transformation with $\hat{x}$ along the Earth-Trojan line of sight and $\hat{y}$ and $\hat{z}$ in the sky-plane as measured from Earth. We set $\hat{z}$ in the direction of the Trojan's sky-plane velocity, while the perpendicular sky-plane direction is set as $\hat{y}$ (see Figure \ref{fig:coordframes}).

The precision to which an asteroid's position can be determined within the sky-plane is dependent upon two factors: (1) the astrometric uncertainty associated with the background star and (2) the network's observing cadence. Occultations do not provide a constraint in the $\hat{x}$ direction for any individual measurement; information from multiple events, however, can be combined to provide constraints in all three directions.

In the $\hat{y}$ direction, the positional uncertainty is equivalent to the occulted star's astrometric uncertainty. In our model, we set all astrometric uncertainties to 20 $\mu$as, corresponding to the projected end-of-mission Gaia astrometric performance for a star in the range V=12-15 mag \citep{Gaia2016, de2005gaia}. This is equivalent to a sky-plane positional uncertainty $\sigma_y\sim75$ m at a distance 5.2 AU; in our simulation, we set uncertainties at each occultation based on the fiducial Earth-Trojan distance at that time.

In the along-track direction within the sky-plane, $\hat{z}$, the asteroid's positional uncertainty is limited by the observing cadence needed to obtain the requisite signal-to-noise ratio (SNR) for the occultation event. We aim for an observing cadence sufficiently high to reach the limits of the Gaia astrometric precision in the along-track direction.

Integrating over visible and near-infrared wavelengths from 400-800 nm, $\sim$4900 photons from a $V=15$ star strike the aperture of a 16-inch telescope each second. To achieve 20 $\mu$as precision at $d=5.2$ AU, we must observe at 225 fps, which results in 21 photons per frame before accounting for inefficiencies in instrumentation. Achievement of this observing cadence is thus permitted by physics and will hinge on the available equipment. Frame rates for CCD and CMOS sensors are limited by the effective number of pixels used by the camera \citep[e.g.][]{el2009cmos}. Because high resolution is not required for our purposes, in which each telescope observes a single star at a given time, we propose that the requisite precision can be attained by binning pixels to meet the appropriate frame rates needed for each star. As a result, we set our uncertainty in the $\hat{z}$ direction to 20 $\mu$as as in the $\hat{y}$ direction.

To quantify the orbital element improvement with improved positional constraints, we employ a Markov Chain Monte Carlo (MCMC) analysis using an affine-invariant ensemble sampler from the \texttt{emcee} Python package \citep{foremanmackey2013}. We consider a set of 5 occultations measured in 1-year intervals, directly reflecting the expected orbital improvement over a 5-year baseline across which Planet Nine's perturbational signal should be observable. We note that we have adopted a conservative occultation measurement rate of 1 yr$^{-1}$, and in practice it therefore should be possible to constrain the orbital elements to even higher precision with more frequent measurements. For comparison purposes, we also consider the case in which 5 occultation measurements are spread evenly across 12 years, corresponding to $\sim 1$ full Trojan orbit.

In each case, we use 100 walkers to sample 6 parameters (the 6 orbital elements), initialized with Gaussian priors where the fiducial orbital elements and associated uncertainties are set as the mean and standard deviation, respectively. At each step of the MCMC, we transform every proposed set of orbital elements into the coordinate frame of each occultation to find where that set of parameters places the asteroid at the measured occultation time. Then, models are accepted or rejected based on the log likelihood function in Equation \ref{eq:loglikelihood}, where $x_n$ and $
\sigma_n$ represent the fiducial asteroid position and associated uncertainty at the $n$th occultation, respectively, while $\mu_n$ represents the model position. We sum over the $\hat{y}$ and $\hat{z}$ directions for each occultation using associated positional uncertainties $\sigma_y$ and $\sigma_z$.

\begin{equation}
    \mathcal{L}(\mu | x, \sigma) = -\frac{1}{2}\sum^{n}_{i=1}\Big(\frac{(x_n - \mu_n)^2}{\sigma_n^2} + \mathrm{ln}(2\pi \sigma_n^2)\Big)
\label{eq:loglikelihood}
\end{equation}

\begin{figure*}
    \centering
    \includegraphics[width=1.0\textwidth]{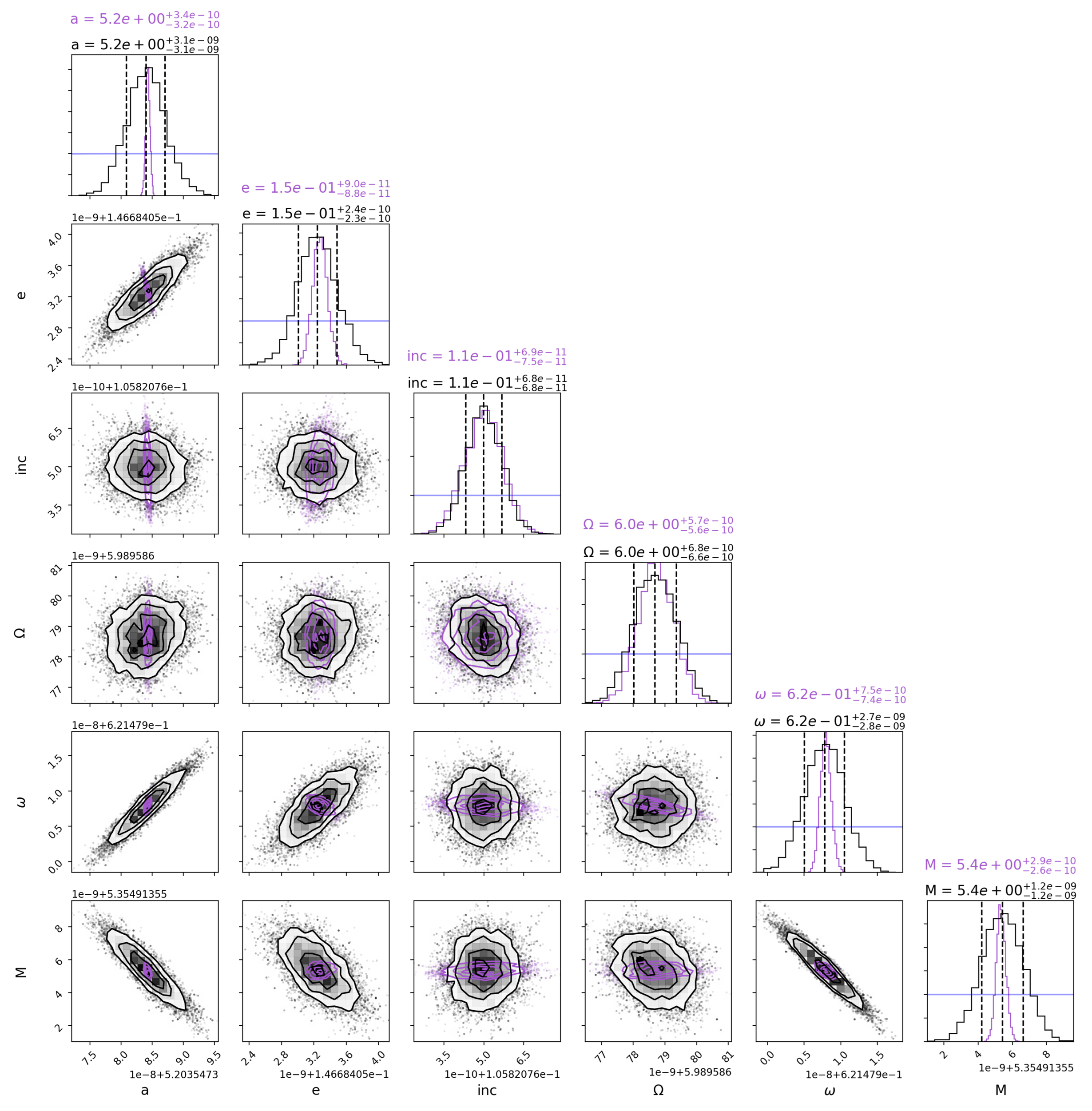}
    \caption{Corner plot showing the distribution of walkers after burn-in for each orbital element after 5 occultations taken with a 1-year spacing. The 16th, 50th, and 84th percentiles of the marginalized posterior probability distribution functions are displayed as dotted lines. In purple, we also include the posteriors where 5 occultations are instead taken at evenly-spaced intervals across the $\sim$12-year Trojan orbit. The blue lines in each histogram display the Gaussian distributions associated with the initial orbital element uncertainties prior to any occultation measurements. The original uncertainties were all improved by over an order of magnitude; as a result, the blue distributions appear flat on this scale. Plot created using the \texttt{corner} Python package \citep{foremanmackey2016}.}
    \label{fig:corner_comparison}
\end{figure*}

We run this process for 1,000 iterations, and we find that the walkers stabilize after $\sim$200 iterations. We discard our first 800 iterations -- well beyond the point of walker stability -- as our ``burn-in" and draw posteriors only from the final 200 iterations. Our results are displayed in Figure \ref{fig:corner_comparison}. The posterior distribution demonstrates that 5 occultation measurements across 5 years can constrain each orbital element to fractional uncertainty $\sigma/\mu \sim 1.5 \times 10^{-9}$ or better. With occultation measurements that are more spread out in time, the width of each posterior shrinks further, reflecting the exquisite precision to which orbits can be determined from occultation measurements over an extended temporal baseline.

\subsection{The Dynamical Effect of Planet Nine vs. the Kuiper Belt}
\label{subsection:P9_v_KB}

By incorporating a two-dimensional model of the Kuiper belt into the EPM2017 (Ephemerides of the Planets and Moon -- 2017 version) planetary ephemerides, \citet{pitjeva2018} found that perturbations to the giant outer solar system planets due to the presence of the Kuiper belt are comparable in magnitude to those induced by a 10$M_{\oplus}$ planet at a solar distance $d=550$ AU. Here, we investigate how the two sources of influence can be disentangled by tracking their respective perturbations on Jupiter's Trojan asteroids.

We examine the change in each orbital element with time due to the presence of an external perturbing body. We complete this study in several steps. First, we compare analytical expressions for the force applied to the Trojan due to a ring and due to a point source (\S\ref{subsubsection:analyt_ext_force}). We then follow the formulation detailed in \citet{murray1999solar} (originally derived in \citet{burns1976elementary}), which derives analytical expressions for the derivative of each orbital element due to an external perturber (\S\ref{subsubsection:analyt_ext_perturb}). Lastly, we simulate the effect of a point source vs. ring perturber in \texttt{REBOUND} to assess net perturbations over an integrated orbit of the Trojan asteroid (\S\ref{subsubsection:num_int_rebound}). Throughout these analyses, we use values listed in Table \ref{tab:P9} and set $\Omega=\omega=0$ and $M=\pi$ to place Planet Nine at apoastron in its orbit.

\subsubsection{Forces Imparted by an External Perturber}
\label{subsubsection:analyt_ext_force}

Evaluation of the instantaneous force imparted by each external perturber illuminates the comparative influences from Planet Nine and from the Kuiper belt, which we approximate as a 2D ring of uniform material. Planet Nine can be treated as a point mass ($m_{P9}$), and thus the acceleration, or force per unit mass imparted upon another point mass a distance $d$ from Planet Nine, is $a_{\mathrm{P9}} = -{Gm_{\mathrm{P9}}}/{d^2}$.

\begin{figure}
    \centering
    \includegraphics[width=0.45\textwidth]{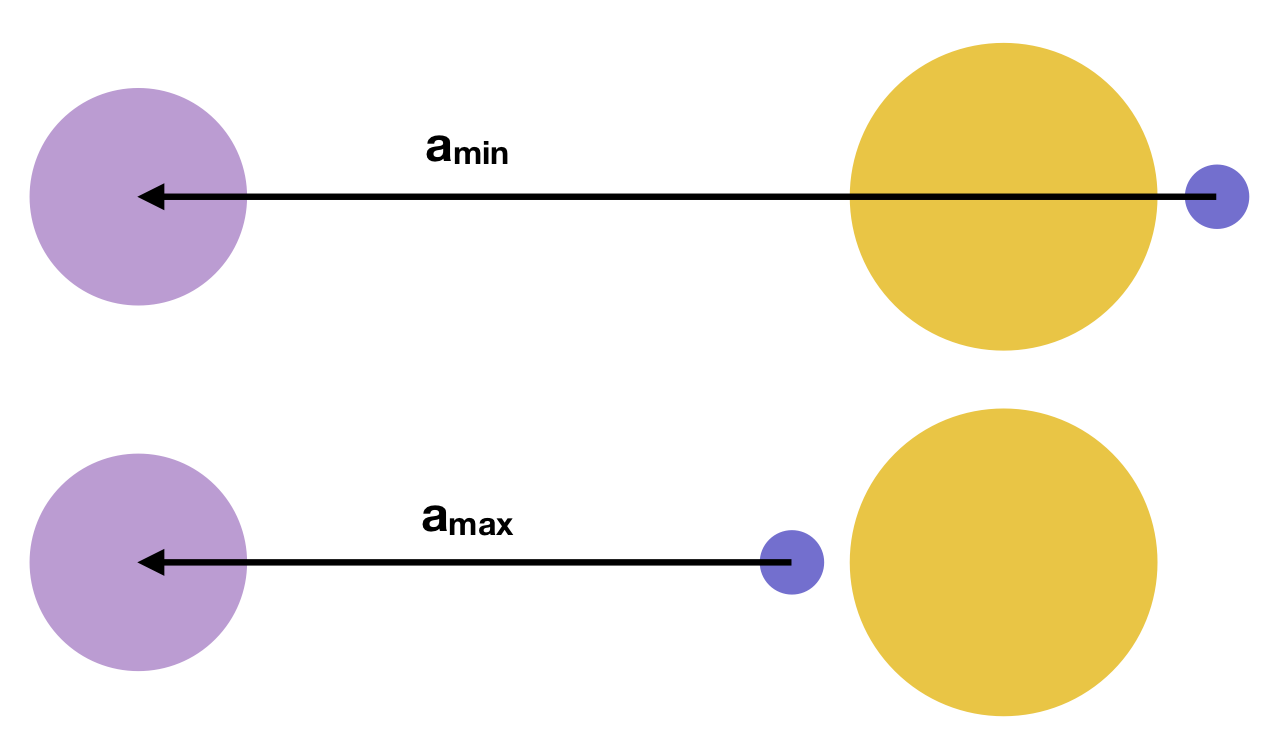}
    \caption{Orbital orientations corresponding to the minimum (top) and maximum (bottom) acceleration imparted upon the test Trojan by Planet Nine. Planet Nine (purple), the Trojan (blue), and the Sun (yellow) are shown with logarithmically scaled relative sizes and distances.}
    \label{fig:P9_KB_alignments}
\end{figure}

We consider two collinear configurations for the Trojan: one in which it is at the farthest point in its orbit from Planet Nine (Figure \ref{fig:P9_KB_alignments}: top), and another in which it is at its close approach to Planet Nine (Figure \ref{fig:P9_KB_alignments}: bottom). The acceleration imparted on the Trojan is $a_{min}$ = 2.01 x 10$^{-11}$ cm/s$^2$ in the former configuration and $a_{max}$ = 2.06 x 10$^{-11}$ cm/s$^2$ in the latter.

\begin{figure}
    \centering
    \includegraphics[width=0.35\textwidth]{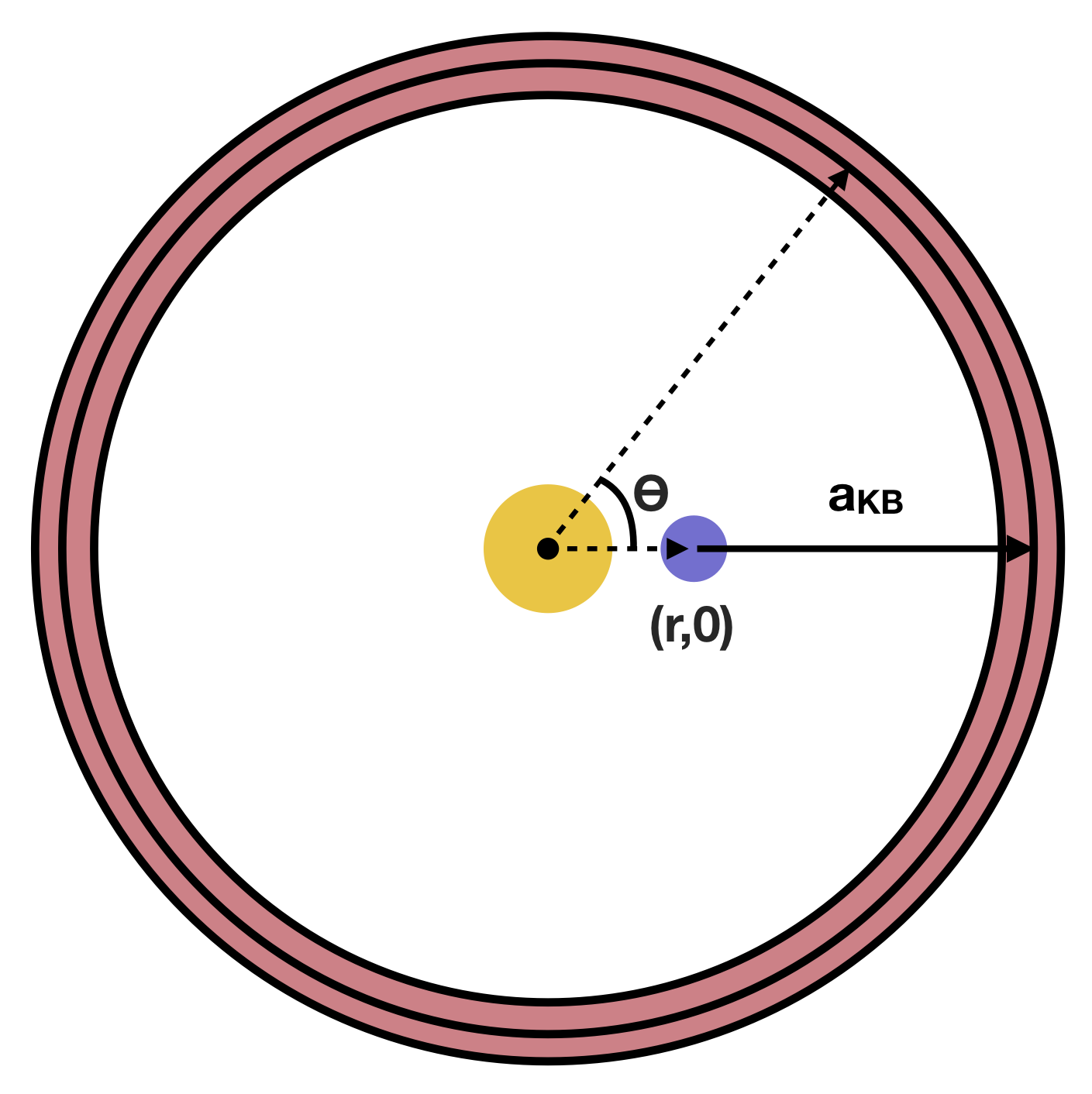}
    \caption{Kuiper belt model used with Equation \ref{eq:ringforce} to integrate acceleration imparted by the ring. We integrate over all $\theta$ to obtain the total $a_{KB}$ = 1.10 x 10$^{-10}$ cm/s$^2$ in the direction of the nearest Kuiper belt edge.}
    \label{fig:KB_model_simplified}
\end{figure}

For comparison, we model the Kuiper belt as a uniform circular ring with radius extending from $R_{min}$ to $R_{max}$, where the angle between Planet Nine and any point in the Kuiper belt is given by $\theta$ (see Figure \ref{fig:KB_model_simplified}). The acceleration imparted upon our test Trojan by the 2D Kuiper belt is given by Equation \ref{eq:ringforce}, where we use the law of cosines to integrate the force per unit mass along the full angular and radial extent of the ring. Here, $\sigma_{KB}$ denotes the 2D mass density of the Kuiper belt. The location of the Trojan is given as $(r,0)$ in polar heliocentric coordinates, where we define the axis representing $\theta=0$ along the line connecting the Sun and the Trojan.

\begin{equation}
   a_{\mathrm{KB}} = -G\sigma_{KB}\int_{R_{min}}^{R_{max}}\int_0^{2\pi}\frac{R(r-R\cos\theta)}{(r^2 + R^2 - 2rR\cos\theta)^{3/2}}d\theta dR
   \label{eq:ringforce}
\end{equation}

We place our test Trojan at $r=5.2$ AU and apply Equation \ref{eq:ringforce} to study the instantaneous perturbation to the Trojan imparted by the Kuiper belt. We obtain surface density $\sigma_{KB}$ by dividing the total mass of the Kuiper belt used in \citet{pitjeva2018},  $m$ = 1.97 $\times$ 10$^{-2}$M$_{\oplus}$, by the full surface area of our 2D disk. We integrate Equation \ref{eq:ringforce} with limits $R_{min}$ = 39.4 AU and $R_{max}$ = 47.8 AU to emulate the disk model from \citet{pitjeva2018}, resulting in $a_{\mathrm{KB}}$ = 1.10 $\times$ 10$^{-10}$ cm/s$^2$. This is approximately an order of magnitude larger than the acceleration obtained from Planet Nine. Due to the symmetry of the Kuiper belt, the magnitude of this acceleration is unaffected by the starting location of the Trojan, though the direction changes throughout the Trojan orbit. This result underscores the findings of \citet{pitjeva2018} and motivates further study to disentangle the perturbational effect of Planet Nine from that of the Kuiper belt.

\subsubsection{Analytical Perturbations to Orbital Elements}
\label{subsubsection:analyt_ext_perturb}
To explore this problem in greater detail, we also consider the effect of each structure on the orbital element evolution of our fiducial Trojan \citep{roy2005book}. We consider a force $F = \bar{R}\hat{r}+\bar{T}\hat{\theta}+\bar{N}\hat{z}$ imparted by a perturber with mass $M_p$, where $\bar{R}$, $\bar{T}$, and $\bar{N}$ are components of force in the $\hat{r}$, $\hat{\theta}$, and $\hat{z}$ cylindrical coordinate directions, respectively. We set all mutual inclinations to zero and thus have $\bar{N}=0$. Zero force applied in the normal direction necessitates that $\dot{I}$ and $\dot{\Omega}$ must be equivalently zero. Expressions for all other orbital element derivatives, following the formulation of \citet{murray1999solar}, are provided in Equations \ref{eq:mu_def}-\ref{eq:taudot}. We report changes in $\tau$, the time of perihelion passage, rather than $M_0$, where the two are related by $M_0=-n\tau$.

\begin{equation}
    \mu = GM_p
    \label{eq:mu_def}
\end{equation}

\begin{equation}
    \dot{a} = 2\frac{a^{3/2}}{\sqrt{\mu(1-e^2)}}[\bar{R}e\sin f + \bar{T}(1+e\cos f)]
    \label{eq:adot}
\end{equation}

\begin{equation}
    \dot{e} = \sqrt{\frac{a(1-e^2)}{\mu}}[\bar{R}\sin f + \bar{T}(\cos f + \cos E)]
    \label{eq:edot}
\end{equation}

\begin{equation}
    \dot{\omega} = \sqrt{\frac{a(1-e^2)}{\mu e^2}}\Big[-\bar{R}\cos f + \bar{T}\sin f\frac{2 + e\cos f}{1 + e\cos f}\Big]
    \label{eq:omegadot}
\end{equation}
    
\begin{equation}
\begin{split}
 \dot{\tau} = \Big[3(\tau-t)\sqrt{\frac{a}{\mu(1-e^2)}}e\sin f \\ + a^2(1-e^2)\mu^{-1}\Big(\frac{-\cos f}{e} + \frac{2}{1+e\cos f} \Big)\Big]\bar{R} \\ + \Big[3(\tau-t)\sqrt{\frac{a}{\mu(1-e^2)}}(1+e\cos f) \\ + a^2 \mu^{-1}(1-e^2)\Big(\frac{\sin f (2 + e\cos f)}{e(1+e\cos f)}\Big)\Big]\bar{T}
\label{eq:taudot}
\end{split}
\end{equation}

We first study the effect of a single point mass perturber to characterize the gravitational impact of Planet Nine. We neglect the tangential and normal components of the force ($\bar{T}=\bar{N}=0$) and focus solely on Planet Nine's radial force. For our purposes, this simplified model is sufficient to obtain an estimate of the expected change in orbital elements over time, since Planet Nine's large predicted separation ensures that its radial force is substantially stronger than forces in the other two reference directions. We calculate $\bar{R}$ using Planet Nine's apoastron distance from the Sun. We report results for Trojan 3794 Sthenelos, again using the orbital elements from Table \ref{tab:3794sthenelos}. The resultant orbital element derivatives are provided in Table \ref{tab:oe_changes}.

\begin{figure}
\centering
\includegraphics[width=0.45\textwidth]{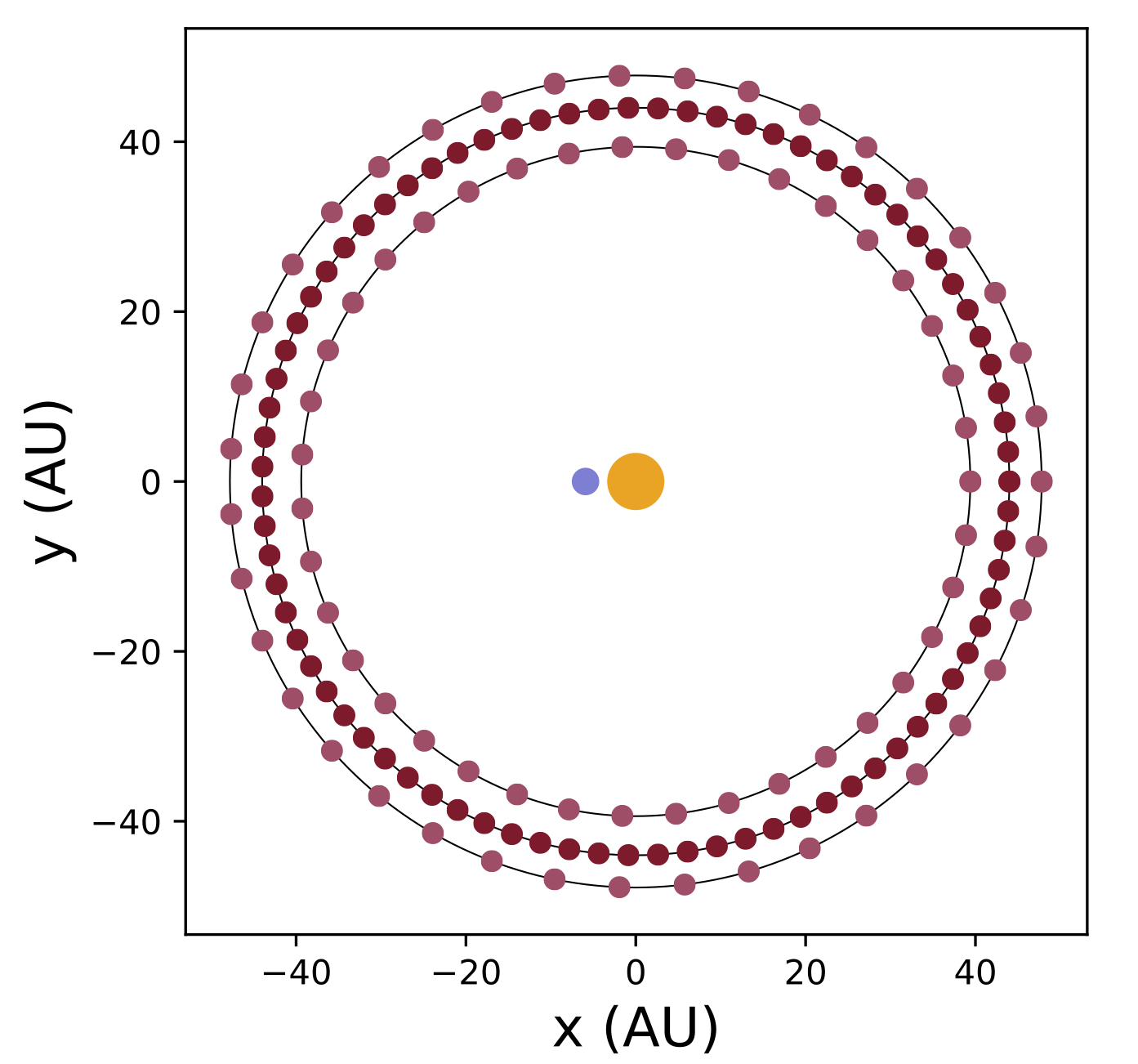}
\caption{Discretized model of the Kuiper belt used throughout \S\ref{subsubsection:analyt_ext_perturb} and \S\ref{subsubsection:num_int_rebound}. A sample Trojan location is shown in blue, while the Sun is in orange (not to scale). There are 80 point masses encompassing the central ring of the Kuiper belt model (dark red), while there are 40 in each the inner and outer rings (light red).}
\label{fig:KB_troj_orbits}
\end{figure}

We also complete this exercise for a circular, 2D ring in the Trojan's orbital plane to obtain the instantaneous change in orbital elements induced by the Kuiper belt. This model includes a tangential force but again has zero normal force. Following the procedure of \citet{pitjeva2018}, our model of the Kuiper belt includes 3 rings, where the inner and outer ring are located at $R_{min}=39.4$ AU and $R_{max}=47.8$ AU, respectively, corresponding to the 3:2 and 2:1 orbital resonances with Neptune. The inner and outer ring each contain 40 discrete, evenly spaced point masses. The central ring contains 80 point masses and is located at $r=44.0$ AU. The total mass of the Kuiper belt, set as 1.97 x 10$^{-2}$ M$_{\oplus}$ \citep{pitjeva2018}, is evenly distributed among these 160 points comprising the Kuiper belt. The model described here is visualized in Figure \ref{fig:KB_troj_orbits}, and the instantaneous changes in orbital elements resulting from the Kuiper Belt are listed in Table \ref{tab:oe_changes}.

\begin{table}[]
    \centering
    \begin{tabular}{c|c|c}
        & Planet Nine & Kuiper Belt \\ \hline
         $\dot{a}$  & -8.50e-10 & 1.69e-7 \\
         $\dot{e}$ & -5.44e-10 & 1.65e-9 \\
         $\dot{\omega}$ & -4.77e-9 & -3.82e-7 \\
         $\dot{\tau}$ & -3.47e-9 & -1.06e-6
    \end{tabular}
    \caption{Instantaneous change in orbital elements induced by Planet Nine and by the Kuiper belt. All values are given in units of the orbital element per year, where $a$ is in AU and $\omega$ is in radians.}
    \label{tab:oe_changes}
\end{table}

The magnitude of the instantaneous change in each orbital element is significantly smaller in the case of Planet Nine than in the case of the Kuiper belt. However, a continuous force in a single direction -- from the effectively static Planet Nine at apoastron -- should have a different effect over time from a ring imparting a symmetric force throughout the Trojan's orbit, where the effect would in part cancel out. We explore this further in \S\ref{subsubsection:num_int_rebound}.

\subsubsection{Numerical Integration with \texttt{REBOUND}}
\label{subsubsection:num_int_rebound}
We next compound the perturbative effects of both Planet Nine and the Kuiper belt on a single Trojan using the \texttt{REBOUND} orbital integrator. We consider two test systems and compare the perturbations upon the Trojan in each: (1) a system including only the Sun, the test Trojan, and Planet Nine, and (2) a system including the Sun, the test Trojan, and the same model of the Kuiper belt described in \S\ref{subsubsection:analyt_ext_perturb} (Figure \ref{fig:KB_troj_orbits}).

All components of each model are treated as point masses, and we integrate over one full orbit of the test Trojan, tracking perturbations from the initial Keplerian orbit over time. For simplicity, we set all inclinations in both models to zero such that all bodies lie within the same orbital plane. We note that our analogous tests which incorporate the inclination of Planet Nine ($\sim$20$\degr$) behave qualitatively in the same way within the plane of perturbations ($\Delta r\times\Delta\phi$) that we consider here.

For consistency, we again provide results for Trojan 3794 Sthenelos, with properties listed in Table \ref{tab:3794sthenelos} with the exception that we set inclination $i=0\degr$. We also run the same \texttt{REBOUND} simulations for several other Trojans of various sizes and confirm that there are no qualitative differences in results. 

\begin{figure}
    \centering
    \includegraphics[width=0.45\textwidth]{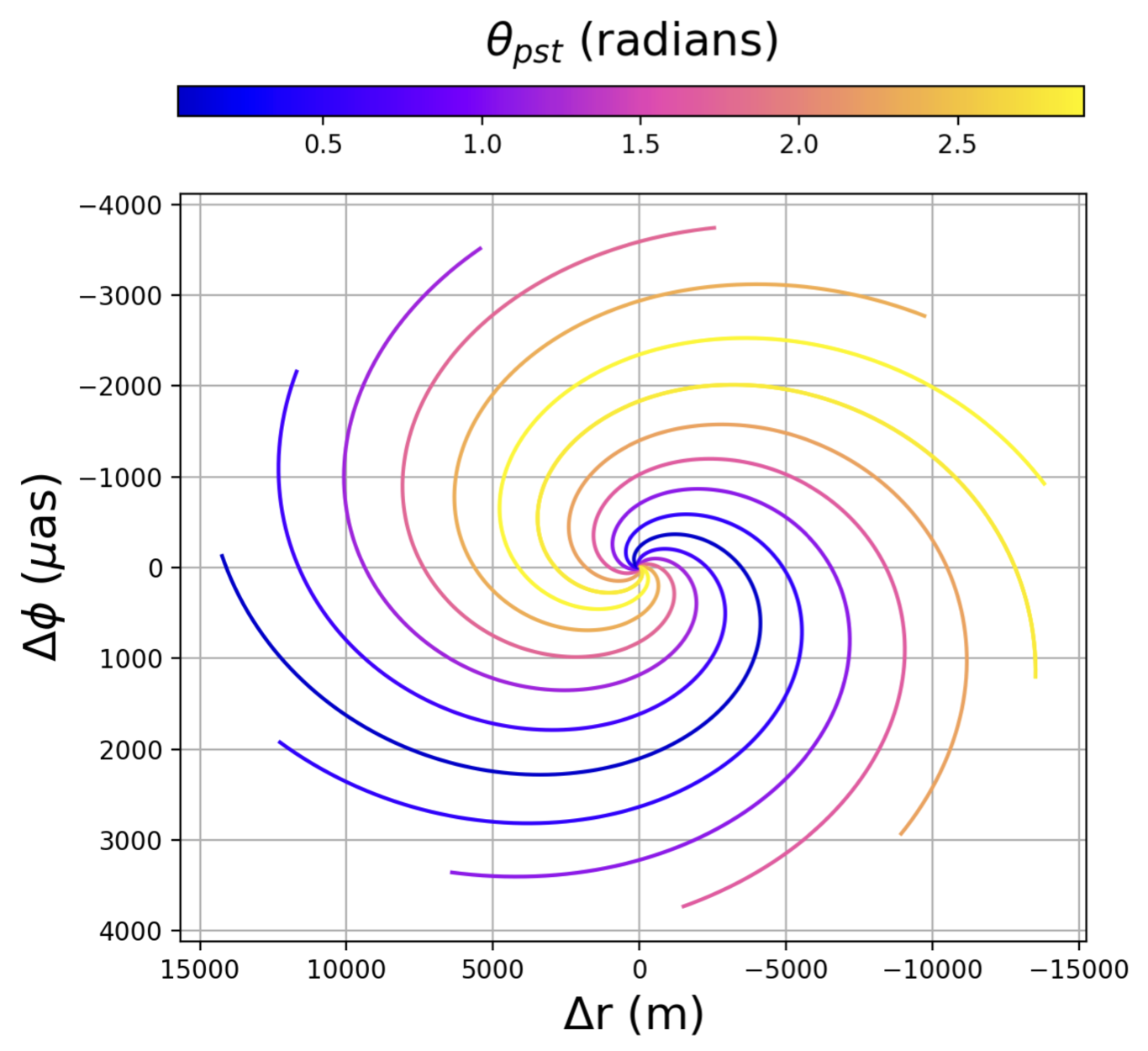}
    \caption{Perturbations induced by Planet Nine, given in spherical coordinates. Each line originates at (0,0) and traces out perturbations over one full Trojan orbit. The color bar indicates different starting points in the Trojan's orbit, where $\theta_{pst}$ provides the Planet Nine-Sun-Trojan angle at the start of the integration. }
    \label{fig:P9_KB_rebound_nosunsub}
\end{figure}

\begin{figure*}
    \centering
    \includegraphics[width=1.0\textwidth]{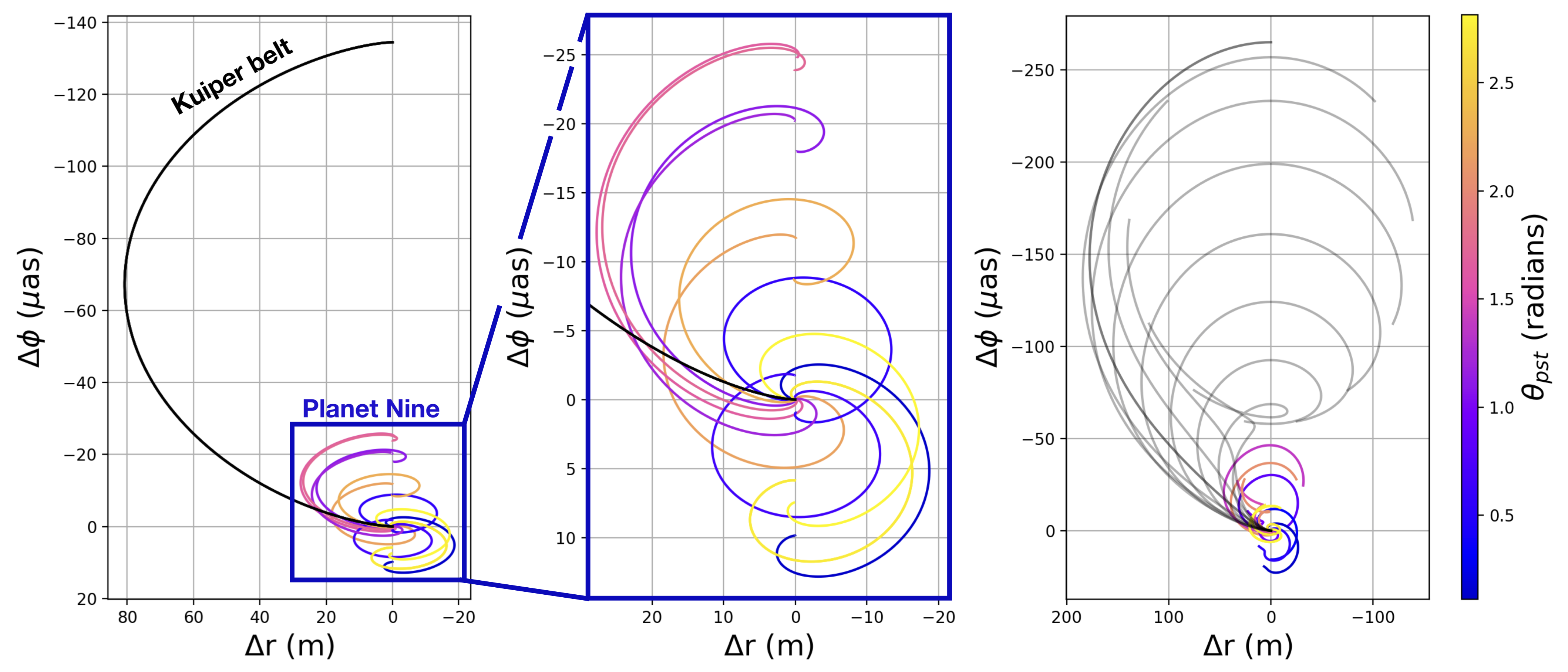}
    \caption{Perturbations induced by Planet Nine (in color) and by the Kuiper belt (in gray), with the Sun's motion subtracted from the system. As in Figure \ref{fig:P9_KB_rebound_nosunsub}, each line originates at (0,0) and traces out perturbations over one full Trojan orbit, while the color scale provides the starting Planet Nine-Sun-Trojan orientation. Results are shown for the case where $e_T=0$ for the Trojan orbit (left and corresponding central panel) and where $e_T=0.147$ from the JPL Small-Body Database (right). In the $e_T=0$ case (left), which is a direct transformation of Figure \ref{fig:P9_KB_rebound_nosunsub}, the Kuiper belt has the same effect irrespective of the initial Trojan location due to the system's symmetry. Incorporating nonzero $e_T$ removes this perturbative symmetry.}
    \label{fig:P9_KB_rebound}
\end{figure*}

The results of our integrations are shown in Figures \ref{fig:P9_KB_rebound_nosunsub} and \ref{fig:P9_KB_rebound}. We use the convention $\Delta\phi$=$\phi_p - \phi_T$ and $\Delta r=r_p - r_T$, where the $p$ subscript refers to the system with a perturber, while the $T$ subscript refers to the system including only the Sun and the Trojan, with no external perturber. Thus, a positive value for $\Delta r$, as is the case for the Kuiper belt where $e_T=0$, indicates that the perturber has pushed the Trojan outwards in its orbit.

In Figure \ref{fig:P9_KB_rebound_nosunsub}, we display the perturbations induced by Planet Nine for a range of initial angles between the Sun-Planet Nine and Sun-Trojan vectors ($\theta_{pst}$). The perturbation induced by the Kuiper belt is not clearly visible on this scale and is thus omitted. At the beginning of each orbit, Planet Nine pushes the Trojan asteroid radially outwards (positive $\Delta r$) for $\theta_{pst} < \pi/2$ and radially inwards (negative $\Delta r$) for $\theta_{pst} > \pi/2$. $\Delta\phi$ is set by the location of the Trojan in its orbit: positive as the Trojan moves towards Planet Nine, and negative as it moves away. Thus, all four quadrants of this space are populated by perturbations, with the perturbative direction set by the Planet Nine-Sun-Trojan orientation.

However, the \textit{differential} perturbation induced by Planet Nine -- the measurable perturbation from within the solar system -- is substantially smaller than the \textit{global} perturbation shown in Figure \ref{fig:P9_KB_rebound_nosunsub}. This is because Figure \ref{fig:P9_KB_rebound_nosunsub} does not incorporate the acceleration also imparted upon the Sun by Planet Nine.

In Figure \ref{fig:P9_KB_rebound}, we subtract the Sun's motion from the system to show the displacement of the Trojan relative to the Sun. To elucidate the behavior of the Trojan in multiple scenarios, we display the perturbations produced in two cases: (1) where the Trojan eccentricity $e_T$ is set to zero (left and central panels in Figure \ref{fig:P9_KB_rebound}), and (2) where we use the fiducial, nonzero Trojan eccentricity extracted from the JPL Small-Body Database (right panel in Figure \ref{fig:P9_KB_rebound}).

For nonzero Trojan eccentricities, the Kuiper belt can produce perturbations in the $+\Delta r$, $-\Delta r$, and $-\Delta\phi$ directions, but never in the $+\Delta\phi$ region of parameter space. In contrast, Planet Nine can produce perturbations in all four quadrants. We obtain similar results when, instead of using the Kuiper belt model in Figure \ref{fig:KB_model_simplified}, we model all but the three most massive TNOs listed in the JPL Small-Body Database (Pluto, Eris, and Makemake). Despite the lack of perfect symmetry in this case, we still find only small ($\sim$5 $\mu$as) deviations into the $+\Delta\phi$ space, suggesting that, in the absence of additional undiscovered Pluto-sized objects, the symmetry of the Kuiper belt is sufficient such that Planet Nine should be gravitationally distinguishable from the Kuiper belt. Given the presence of an additional Pluto-sized perturber in the Kuiper belt, this network would provide strong evidence for its existence and constraints upon its location -- an interesting result in itself.

\begin{figure}
    \centering
    \includegraphics[width=0.45\textwidth]{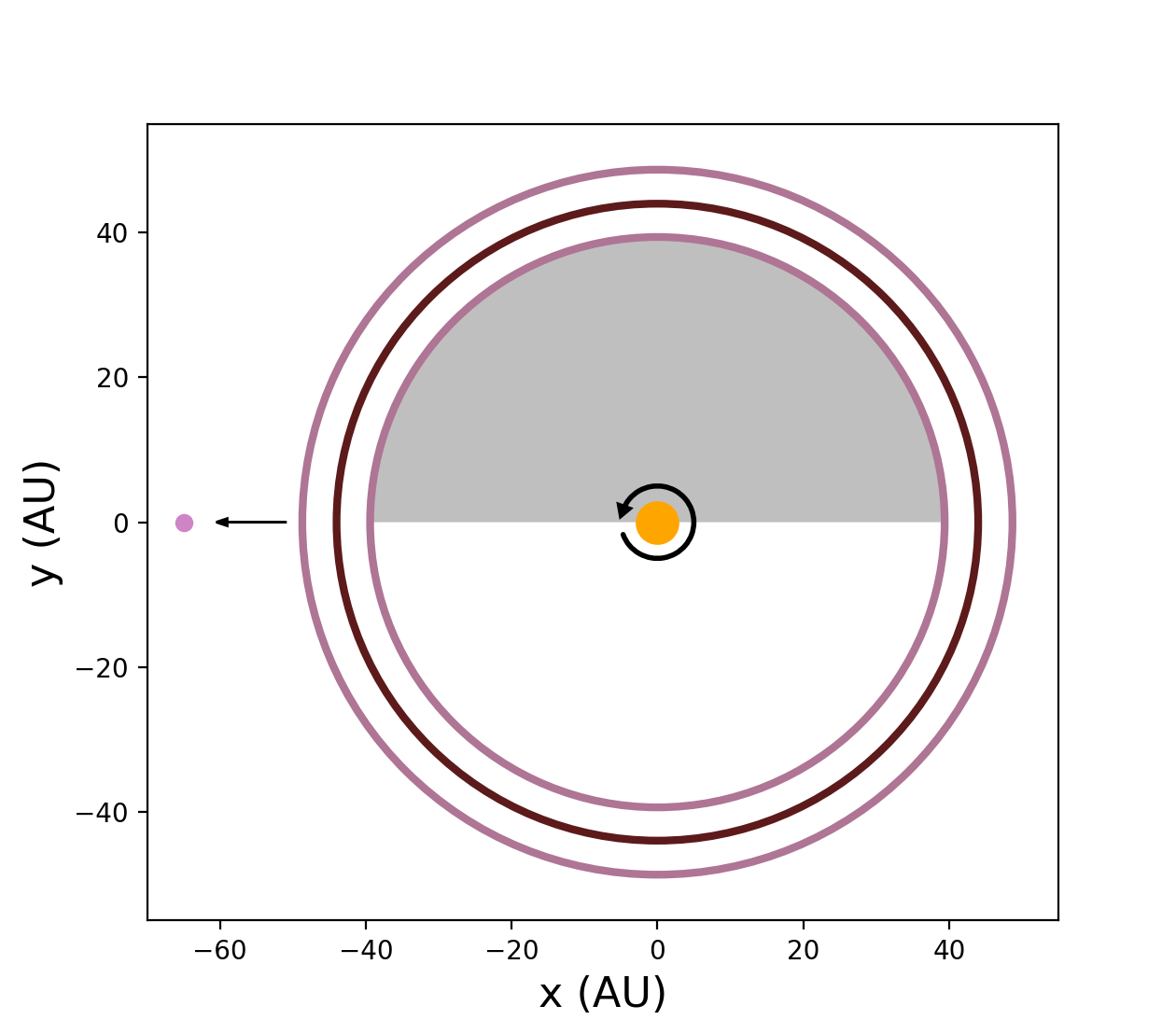}
    \caption{By observing Trojans only in the half of the sky shaded in gray, where $\Delta\phi$ from Planet Nine is positive, it is possible to differentiate perturbations from Planet Nine from those of the symmetric Kuiper belt, which always correspond to negative $\Delta\phi$. The direction of the Trojan orbit and of Planet Nine's position to the left are displayed for orientation, while the Kuiper belt is represented by 3 concentric rings.}
    \label{fig:P9_where_obs}
\end{figure}

It is possible to distinguish the effects of Planet Nine and the Kuiper belt, therefore, by studying occultations of the Trojans in regions of the sky which correspond to positive $\Delta\phi$ for a Planet Nine perturber, but which still result in a negative $\Delta\phi$ from the Kuiper belt. These include regions in which the Trojan is moving towards Planet Nine in its orbit and thus is pulled further forward by the perturber, as illustrated in Figure \ref{fig:P9_where_obs}. The perturbational signature of the putative Planet Nine will deviate most from the noise produced by the Kuiper belt during the half of the Trojan orbit in which perturbations are in the $+\Delta\phi$ direction. The peak of this $+\Delta\phi$ deviation occurs when $\theta_{pst}=\pi/2$ and the Trojan is moving towards Planet Nine. As a result, the angular orientation of Planet Nine can be extracted upon detection of the predicted perturbation from a distribution of Trojans.

The perturbation size detectable by our network is limited by the angular precision of the occulted star ($\omega$=12-25 $\mu$as at $V=15$ from the Gaia catalog). This is much more precise than standard astrometric measurements, and even astrometry with HST typically results in uncertainties of order $\sim$1 mas or greater \citep[e.g.][]{bellini2011astrometry, porter2018} and can only reach precision $\sim0.1$ mas for the very brightest targets \citep{benedict2002astrometry}. Thus, the extremely high precision necessary to observe Planet Nine's perturbational effect can be decisively obtained through the occultation method. 

By measuring a large number of Trojan positions over time and searching for this $+\Delta\phi$ signature of a point perturber, we can extract the Planet Nine signal and separate it from that of the Kuiper belt. For an individual small Trojan, we have shown that ephemeris refinement from successive occultations through the course of a full orbit generates a final positional accuracy of order $\delta {\bf x}=75{\rm \,m}$. Figure \ref{fig:P9_KB_rebound}, on the other hand, suggests that systematic deflections on the Trojan orbit due to Planet Nine are of order 25 meters, corresponding to a detection of $\sigma_i={1/3}\, \sigma$ significance. As a consequence, the tracking of a single Trojan is no more useful for detecting Planet Nine than was the Cassini telemetry. With a large occultation network, however, thousands of bodies can be monitored, with detection significance rising to $\sigma_{\rm final}=\sigma_{i}\sqrt{N}$, where $N$ is the number of Trojans whose orbits have been improved to the DR2-enabled precision. As an example, refinement of $N\sim225$ Trojan orbits would generate $\sigma_{\rm final}=5\sigma$ detection significance.

\section{Additional Applications}
\label{section:applications}

The construction of the proposed cISP occultation network would have considerable potential not only in the search for Planet Nine, but also in a wide variety of other applications. We showed that regular stellar occultation measurements would deliver unprecedented constraints on the orbital ephemerides of Trojan asteroids. This concept can be extended to various other small-body populations within the solar system, including near-Earth objects, main-belt asteroids, and TNOs, delivering unprecedented constraints on the distribution of minor body shapes and sizes within the solar system. Such applications can be utilized in a variety of ways.

In this section, we describe a selection of additional scientific use cases for the network. We focus on four sample applications: (1) synergies with planned NASA missions (\S\ref{subsection:NASA_synergies}), (2) investigating the origins of Jovian and Neptunian Trojans (\S\ref{subsection:trojans}), (3) using TNO occultations to develop an acceleration map of the outer solar system (\S\ref{subsection:TNOs}), and (4) independently verifying asteroid diameters throughout the solar system (\S\ref{subsection:diameters}).

\subsection{Synergies with Future NASA Missions}
\label{subsection:NASA_synergies}

\subsubsection{The Lucy Mission}
As an added benefit of the search for Planet Nine, improved constraints on the orbital ephemerides of the Jovian Trojan population may assist NASA's \textit{Lucy} Mission, which is planned for launch in October 2021 and will fly by seven asteroids -- including six Jovian Trojans -- from 2025-2033 \citep{levison2016lucy, levison2017lucy}. Improving the positional accuracy of the Jovian Trojans can potentially enhance the safety of the spacecraft, conserve fuel, and improve trajectory planning. Occultation measurements can also probe for rings and other opaque structures in the Trojan systems prior to spacecraft arrival.

Moreover, a combination of \textit{in situ} range observations from the spacecraft and near-simultaneous Jovian Trojan positional measurements obtained via occultations would generate fully three-dimensional positions of extraordinary accuracy for the handful of Jovian Trojans studied in-depth by \textit{Lucy}. These exquisite 3D positional measurements would be highly complementary to the larger statistical sample that we propose studying throughout this work.

\subsubsection{The Psyche Mission}

The \textit{Psyche} Mission is planned for launch in 2022 and will arrive at Main Belt asteroid (16) Psyche in 2026, where it will orbit for 21 months \citep{oh2016psyche, lord2017psyche}. Equipped with an X-band telecommunications system and a multi-spectral imager, the mission spacecraft will map the gravity field of Psyche while also characterizing its topography \citep{hart2018overview}. 

By concurrently measuring occultations of Psyche as it is studied by the orbiting spacecraft, it will be possible to not only constrain the asteroid's position to exquisitely high precision (as for the Trojans studied by \textit{Lucy}) but also to simultaneously obtain 2D projected spatial maps of the asteroid from multiple viewing points. Synergistic measurements from both an occultation network and the orbiting spacecraft will provide extremely detailed constraints on the asteroid's three-dimensional topography.

\subsection{Origins of the Jovian \& Neptunian Trojans}
\label{subsection:trojans}

Trojan asteroids, introduced in the Jovian context in Section \ref{subsection:casetidalacc}, are more generally defined as small bodies librating at the L4 and L5 Lagrange points of a planet-star system in a 1:1 mean motion resonance, first theorized by \citet{lagrange1873oeuvres} as a stable solution to the restricted three-body problem. \citet{wolf1906} discovered the first known Trojan asteroid at Jupiter's L4 point, and the vast majority of Trojan asteroids discovered since then are also located in the Jovian system.

More recently, a population of Trojans was also found in the Neptunian system \citep{chiang2003, sheppard2006thick, sheppard2010a, sheppard2010b, becker2008}, with a comparable number of asteroids at each Lagrange point and similar dynamics in both populations. Previous works have explored possible capture mechanisms for the Jovian and Neptunian Trojans \citep[e.g.][]{marzari2002origin, morbidelli2005chaotic, sheppard2006thick, lykawka2009origin}, where most of the recent models incorporate capture from the same population that now comprises the modern-day Kuiper belt, as in the Nice model \citep{tsiganis2005}. However, the exact capture mechanism of these two populations remains to be definitively established. 

\citet{sheppard2006thick} and later \citet{jewitt2018} observed that the color distribution of the Jovian Trojans is statistically indistinguishable from that of the Neptunian Trojans -- consistent with a shared origin for both populations. However, \citet{jewitt2018} found that this distribution does not match that of any dynamical sub-population within the Kuiper belt, where these sub-populations are defined in \citet{peixinho2015}. This discrepancy refutes the hypothesis that the two Trojan populations were captured as scattered KBOs, raising the question of where else the Trojans may have instead originated.

A robust comparison of asteroid size distributions would serve as another useful distinguishing factor, since the parent population should have a similar size distribution to that of the captured Trojans. The $H$ magnitude distribution of the large Jovian Trojans, which is closely tied to the radius distribution, has been measured on several occasions and matches a steep power law with index $\alpha$ ($\alpha = 0.9\pm0.2$ \citep{jewitt2000}, $\alpha = 1.0\pm0.2$ \citep{fraser2014}, $\alpha = 0.91\pm^{+0.19}_{-0.16}$ \citep{wong2015color}); however, the paucity of bright Trojans limits the accuracy to which $\alpha$ can be determined. The translation of an $H$ magnitude distribution to a radius distribution also requires assumptions about the albedos of the Trojans, which could be circumvented through direct size observations using stellar occultations. Currently, the ranges of asteroid sizes measured for the Trojan and Kuiper belt populations barely overlap -- thus, precise constraints on both $\alpha$ and the size distribution of small KBOs would enable a more convincing direct comparison.

\subsection{An Acceleration Map of the Outer Solar System}
\label{subsection:TNOs}

Another appeal of this network lies within its ability to pinpoint the positions of distant minor planets, such as TNOs, to high precision. While challenging, stellar occultation measurements of TNOs have been successfully demonstrated on several occasions \citep{ortiz2017size, muller2018haumea, sickafoose2019stellar} and have become increasingly accessible with the advent of Gaia DR2. Deep sky surveys such as the Dark Energy Survey \citep{dark2005dark} can also provide astrometric measurements of TNOs with sufficient accuracy to reliably predict upcoming occultation events \citep{banda2018astrometry}. By observing TNO occultations over time, it is possible to constrain the allowed parameter space for undiscovered minor planets in the outer solar system, leading to an effective ``acceleration map." As new extreme TNOs with $a>250$ AU are discovered, such a map will contextualize the distribution of known objects and constrain their abundance. 

Furthermore, occultation measurements provide the added benefit of determining TNO shapes and sizes. Stellar occultation measurements are currently the only ground-based observational technique capable of constraining the geometry of TNOs to kilometer accuracy. However, as described in Appendix \ref{section:diffraction}, the Fresnel diffraction limit becomes important for small TNOs and must be taken into consideration when planning TNO occultation measurements.

\subsection{Asteroid Diameter Verification}
\label{subsection:diameters}

Occultation measurements provide an independent test to verify previously obtained asteroid diameters, as well as tighter constraints to decrease the uncertainties in reported values. A combination of measurements from the WISE \citep{wright2010} and NEOWISE \citep{mainzer2011} infrared surveys has been modeled using the Near-Earth Asteroid Thermal Model (NEATM; \citet{harris1998thermal}), resulting in a database of measured diameters for $\sim$150,000 asteroids \citep{mainzer2016neowise}. The methods and underlying assumptions used by the NEOWISE team to obtain these asteroid diameters were recently challenged \citep{myhrvold2018asteroid, myhrvold2018empirical, myhrvold2018response}, with a corresponding rebuttal in \citet{wright2018response}. Independent measurements obtained through separate methods can improve confidence in reported values obtained through thermal modeling, and this could serve as an external test of the accuracy of previously measured values.


\section{Camera \& Telescope Selection}
\label{section:camera_telescopes}

The availability of funding will play a deciding role in the specific magnitude limits of this network, and we envision that financial support for the network would potentially leverage private-public partnerships. In this section, we discuss desired properties for the network and provide cost estimates based on several currently operating camera systems optimized for occultation observations. While we focus on hardware configurations and their associated expenses, we note that operating costs are nontrivial and would play a material role in the final project budget.

A large fraction of the total construction cost will be encompassed by camera expenses. We require a camera capable of observing at a high cadence with minimal dead time between frames in order to obtain the highest possible time resolution and corresponding spatial resolution. The field of view should ideally encompass at least 10$'$ to include a sufficient number of comparison stars for calibration. To push our magnitude limit and timing cadence as low as possible, we benefit from a camera with the highest possible quantum efficiency, maximizing the percentage of incoming photons that are converted to detected electrons. For faint objects occulting faint stars, the dominant noise source will likely be read noise; as a result, an optimal detector will minimize this quantity.

For comparison, we consider the hardware setup used by the New Horizons team in preparation for the MU69 flyby. This configuration was sufficient to obtain five chords of varying lengths across MU69, all consistent with the bi-lobed structure observed during the flyby. To measure an occultation over a $V\sim15$ star, the New Horizons team used a set of Sky-Watcher 16'' truss-tube Dobsonian reflectors ($\sim$\$2000 each) paired with QHY174M-GPS CCD cameras ($\sim$\$1240 each with cooling).\footnote{\url{https://www.skyandtelescope.com/astronomy-news/solar-system/observers-track-new-horizons-next-target/}} These CCD cameras have $78\%$ quantum efficiency and can run between 138-490 fps depending on the user's pixel binning. Two thousand sets of this telescope-CCD pairing would amount to \$6.5 million. While this particular configuration is not endowed with robotic controls, switching the telescope model for an automated version with the same photon collecting area should provide similar results. Potential alternatives include 16-inch automated telescopes such as the Meade LX600-ACF 16" (f/8) Cassegrain telescope (\$19,000 per unit, including tripod mount) or the Meade LX200-ACF 16" (f/10) computerized telescope (\$13,500 per unit, without mount).

We also consider the possibility of adapting previous iterations of optimized occultation camera systems for our purposes. We focus on four recent systems: the Portable Occultation, Eclipse, and Transit System \citep[POETS;][]{souza2006}; the Portable High-Speed Occultation Telescope \citep[PHOT;][]{young2011}; the Portable Instrument for Capturing Occultations \citep[PICO;][]{lockhart2010pico} and the Astronomical Digital Video occultation observing and recording System \citep[ADVS;][]{barry2015}. We provide a brief overview of each of these CCD systems here, and we refer the reader to Table 3 in \citet{barry2015} for a detailed comparison of their properties.

POETS was developed from 2005-2006 as a portable camera system small enough to fit into two standard carry-on luggage bags. With frame rate $\sim$35 frames per second (fps), high quantum efficiency ($\geq$90\% from 500-700nm), and low read noise ($<$1-49 e$^-$ at 10 MHz), POETS has been used for a wide variety of observations ranging from exoplanet light curve observations \citep{winn2009transit, adams2011twenty} to studies of Pluto's atmosphere \citep{bosh2015state} and total solar eclipse observations \citep{pasachoff2009scientific}. POETS uses a frame-transfer electron-multiplying CCD for which dead time between frames is nearly zero and limited only by the time required to shift accumulated charge to the storage region of the CCD (typically on the order of ms). Frame-transfer CCDs allow for high-cadence observations with high time resolution, making them suitable for occultation measurements. They are, however, relatively expensive, and thus we do not explore this option further.

The PHOT system has very similar specifications to those of POETS and has been used to study several stellar occultations of Pluto and Charon \citep{young2008vertical, young2010results, young2011occultations, olkin2014pluto}. Like POETS, the PHOT system also uses an expensive frame-transfer CCD, so we do not discuss it in further depth here.

The PICO system, which is significantly less expensive than POETS and PHOT (\$5,000 vs. \$38,000 and \$30,000, respectively)  has read noise $\sim$13.6 e$^-$ at 2.8 MHz and lower quantum efficiency peaking at 58\%. It does not employ a frame-transfer CCD and has significant inter-frame dead time ranging from 0.65-0.76 seconds per frame, resulting in total observing cadence 1.63 fps for a sample observation of KBO 762 Pulcova \citep{lockhart2010pico}. With average shadow speeds $\sim$17 km/s for Jovian Trojans and $\sim$25 km/s for KBOs, this cadence limits occultation detections to objects with minimum diameter 55 km and 82 km, respectively, where at least two in-occultation observations are required for an individual detection. Though this system would be sufficient to observe large bodies, we desire a shorter observing cadence to probe a large number of Jovian Trojans, many of which are significantly smaller than this diameter limit.

The final and most recently deployed system that we discuss here is the ADVS, which uses the GX-FW-28S5M-C Grasshopper Express CCD camera with frame rate 26 fps, quantum efficiency 68\%, read noise 9.6 e$^-$, and inter-frame dead time $<$2 ms. The full system costs \$4,500, and, with frame rate 26 fps, it would be possible to probe objects with diameters down to $\sim$1.3 km for Jovian Trojans and $\sim$1.9 km for KBOs. While this performance is sufficient to observe Trojans down to our size limit, a camera with a higher frame rate would be necessary to reach the limits of the Gaia astrometric precision for $V=15$ stars (225 fps or better to achieve 20 $\mu$as precision at 5.2 AU).

Beyond these previously implemented configurations, which all employ CCD systems, thermoelectrically cooled CMOS cameras provide another promising alternative for low-cost, low-read noise photometry at a high frame rate, with several versions currently available reaching $\geq$225 fps. Notably, two sets of a CMOS camera/11" telescope configuration (\$16,000 for each set) were recently used to observe an occultation of a 1.3 km KBO, where both cameras successfully observed the event \citep{arimatsu2019}. An in-depth exploration of the CMOS option would be warranted upon implementation of this network. However, we choose not to explore this option further here due to the substantial technological improvements that will likely be made prior to the time of network construction.

Because the cISP network adapts only already-existing communication towers, infrastructure expenses for this project would be minimized. In addition to the telescope/camera systems themselves, automated telescope covers would be necessary to protect and efficiently deploy all units. These expenses will constitute a small fraction of the budget by comparison with the driving costs of the telescope/camera systems employed throughout the grid. We estimate that, for bulk purchases, telescope housing and similar construction expenses will amount to $\sim$\$1500 per telescope, or \$3 million in total for all 2000 telescopes.

\section{Network Multiplicity}
\label{section:multiplicity}

We have presented this network under the assumption that one telescope would be placed at \textit{every} location throughout the cISP network; here, we also consider the case in which telescopes are placed not at all possible sites, but only at some subset. For the following reasons, we do not explore this possibility outside of the context of the proposed network site distribution:

\begin{enumerate}
\item Finding and setting up new sites that are not part of a pre-existing framework would substantially increase infrastructure expenses. As a result, the starting overhead cost of the network would be much higher when including sites that have not been pre-specified.
\item The quasi-random spread of telescopes across the country decreases the likelihood that the full network will ever be down at the same time due to poor weather. Thus, a more compact setup of telescopes is disfavored.
\end{enumerate}

We acknowledge that an analogous network with a different set of sites and similar quasi-random spread of telescopes could achieve the same goals outlined here; however, an exploration of alternative network configurations is beyond the scope of this work.

A reduction in the number of network telescopes could substantially reduce the price of the network; in fact, it is likely that certain sites may not be amenable to astronomical observations due to light pollution. However, we argue that most, if not all, sites should be maintained within the network in order to maximize its potential for a range of scientific cases.

A smaller number of sites incorporated into the network would increase the average latitudinal gap between telescopes, meaning that occultations would more frequently slip through gaps in the network. This disadvantage would most strongly affect small ($\sim$km-size) asteroids, drastically reducing the total number of consistently observable occultation events. Individual asteroids would also become more difficult to regularly track due to this reduced efficiency.

Beyond increasing the likelihood that any \textit{individual} occultation event is observable, a large number of telescopes also increases the likelihood that a single occultation event may be observable at multiple sites. Observations of a single occultation event by multiple telescopes provide tighter constraints upon the size of an individual object while also improving confidence in low-SNR occultation measurements. Thus, auxiliary science cases, such as constraining the size distribution of asteroid populations, benefit substantially from the use of a large number of telescopes. Since we propose this network not only as a way to find Planet Nine, but also as a tool to better characterize the distribution of small bodies in the solar system, we choose to present it as a whole such that, upon implementation, it would be as useful as possible for a range of applications.

\section{Conclusions}
\label{section:conclusions}
In this paper, we discussed the scientific merits of a national network of telescopes designed to observe occultations of solar system asteroids over background stars. We focused on an in-depth analysis of the network's capabilities as a probe for the proposed Planet Nine, where we use Jovian Trojan asteroids as trackers of the tidal differential acceleration imparted by Planet Nine. In summary, we find the following:

\begin{itemize}
\item Each Jovian Trojan asteroid occults a  $V\leq15$ Gaia star with a median rate of 7.42 occultations per Trojan per year, corresponding to $\sim$1.0$\times 10^5$ total observable events per year when accounting for the day/night cycle. These rates require both solar angle below $-8\degr$ and airmass $z<2$ while the Trojan is in the sky. In practice, some Trojans will occult background stars more frequently, while others will occult less frequently. This overall rate is sufficient to track many individual Trojans over a timescale of a few years in search for the gravitational signature of Planet Nine.
\item We quantify the improvement in orbital element uncertainties over several occultations, showing that the precise positional constraints from occultation measurements directly translate to exquisite constraints on the asteroid's orbital elements. We find that 5 occultations observed over 5 years constrain each orbital element to fractional uncertainty $\sigma/\mu \sim 1.5 \times 10^{-9}$ or better, while 5 observations evenly spaced over a full $\sim12$-year Trojan orbit provide even tighter constraints.
\item It is possible to distinguish the gravitational perturbing effects of an encircling, (nearly) symmetrically distributed Kuiper belt from that of Planet Nine by strategically observing Jovian Trojans within the half of their orbit in which they are moving towards the effectively stationary location of Planet Nine. This is due to the differing gravitational signature of a circular, approximately symmetric system -- the Kuiper belt -- by comparison to a point mass -- Planet Nine.
    
\end{itemize}


We also outlined several further science cases for the network. In summary, we find that occultation measurements by the proposed network can (1) provide complementary observations to support and supplement the NASA \textit{Lucy} and \textit{Psyche} missions, (2) deliver direct, robust measurements to compare the Jovian and Neptunian Trojan size distributions as a test of solar system formation models, (3) permit the development of an outer solar system acceleration map through high-precision measurements of TNOs, and (4) provide an independent test of previously measured asteroid diameters.

Throughout this work, we focus on one configuration for a telescope grid in which all locations are tied to sites along the cISP low-latency communications network. Under the assumption that the latitudinal distribution of telescopes is not substantially altered, the same concept of a large-scale occultation network could, in theory, be implemented with deviations from these specifications without significant changes to the scientific case for the network. In particular, reducing the size of the largest latitudinal gaps present in the current telescope distribution -- either by adding/adjusting telescope sites within the current configuration or by using an altogether separate configuration to implement the same concept -- would improve the rate of successful occultation measurements for primarily East-West occulting asteroids.

Similar projects may also consider leveraging the availability of university observatories and small telescopes worldwide. Though this would lead to a non-uniformity of telescope systematics, incorporating these facilities would expand the baseline of possible observations. In any of these forms, a large-scale occultation network would serve as a powerful and timely tool to utilize synergies between Gaia and LSST, demonstrating the remarkable potential of small telescopes in the upcoming era of astronomy.

\section*{Acknowledgements}
\acknowledgments

We are grateful to Konstantin Batygin, Andy Szymkowiak, and Frank van den Bosch for helpful comments. We also thank the anonymous referee for constructive feedback that led to a substantial improvement of the manuscript. M.R. is supported by the National Science Foundation Graduate Research Fellowship Program under Grant Number DGE-1752134. This material is also based upon work supported by the National Aeronautics and Space Administration through the NASA Astrobiology Institute under Cooperative Agreement Notice NNH13ZDA017C issued through the Science Mission Directorate. We acknowledge support from the NASA Astrobiology Institute through a cooperative agreement between NASA Ames Research Center and Yale University. Simulations in this paper made use of the \texttt{REBOUND} code which can be downloaded freely at \url{http://github.com/hannorein/rebound}. This research also made use of the \texttt{numpy} \citep{oliphant2006guide, walt2011numpy}, \texttt{matplotlib} \citep{hunter2007matplotlib}, \texttt{astropy} \citep{astropy:2013, astropy:2018}, \texttt{pandas} \citep{mckinney2010data}, \texttt{emcee} \citep{foremanmackey2013}, \texttt{corner} \citep{foremanmackey2016}, and \texttt{Altair} \citep{vanderplas2018altair} Python packages.

\appendix
\section{Assigning Jovian Trojan Diameters}
\label{section:troj_diameters}

Where diameters are not listed in the JPL Small-Body Database, we convert the reported absolute magnitudes ($H$) to diameters using Equation \ref{eq:H_to_rad}, derived from Equations \ref{eq:flux_mag_relation} and \ref{eq:flux_intensity_relation} where $F$ refers to flux, $m$ is magnitude, $I$ is intensity, and $\Omega$ is the solid angle over which $I$ is collected.

\begin{equation}
D = \frac{2(1 \,\mathrm{AU})}{\sqrt{p_v}} 10^{\frac{1}{5}(m_{\odot}-H)}
\label{eq:H_to_rad}
\end{equation}

\begin{equation}
\frac{F_1}{F_2} = 10^{-\frac{2}{5}(m_1 - m_2)}
\label{eq:flux_mag_relation}
\end{equation}

\begin{equation}
F = I\Omega
\label{eq:flux_intensity_relation}
\end{equation}

In Equation \ref{eq:H_to_rad}, $m_{\odot}=-26.74$ is the absolute magnitude of the Sun, and $H$ and $p_v$ are the absolute magnitude and geometric albedo of the small body, respectively. Albedo values have not been measured for the vast majority of Jovian Trojans; however, \citet{fernandez2003} observed a mean albedo of $0.056\pm0.003$ for 32 Jovian Trojans with diameter $D\geq25$ km. Later, \citet{fernandez2009} reported a mean albedo $0.121\pm0.003$ for a separate sample of 44 small (5 km$<D<24$ km) Jovian Trojans. Accordingly, we assign albedo $a=0.056$ to all Jovian Trojans and, for those objects which return $D\leq25$ km, we reassign albedo $a=0.121$.



\section{Diffraction Effects}
\label{section:diffraction}

The Fresnel diffraction limit must be taken into consideration for observations of small-body populations employing stellar occultations. This limit quantifies the size scale of occulting objects below which diffraction effects become non-negligible. The Fresnel diffraction limit is given in Equation \ref{eq:fresnel}, where $\lambda$ is the wavelength of observation and $d$ is the distance from the telescope to the occulting object.

\begin{equation}
\label{eq:fresnel}
    F = \sqrt{\lambda d/2}
\end{equation}

Objects with radius $r\sim F$ display wavelength-dependent diffraction fringes that are irregular for objects with rough edges. At visible wavelengths ($\lambda=700$ nm), the Fresnel scale for Jovian Trojans at distance $d=5$ AU is $F=0.5$ km, while $F=1.5$ km for TNOs at a distance $d=45$ AU. We anticipate that the Fresnel limit may need to be taken into account in the case of our smallest targets, particularly in the case of TNOs and other distant targets with relatively large $F$. Fresnel-Kirchoff diffraction theory can be applied in such cases to model objects with $r\sim F$.

\section{(Non)-Detectability of Dark Matter}
When a massive body moves through a sea of less massive bodies, it experiences a loss of momentum and kinetic energy through a process called dynamical friction, which produces a trailing wake of concentrated density behind the massive body. \citet{hernandez2019detailed} recently proposed that the sun's motion through the Milky Way should produce a wake of dark matter particles detectable through gravitational perturbations within the solar system. For estimated dark matter particle dispersion $\sigma$ = 200 km/s and density $\rho_0 = 0.01 M_{\odot}$ pc$^{-3}$ \citep{read2014local}, we use Equation 2 of \citet{hernandez2019detailed} to find that the tidal acceleration of the dark matter wake at 10 AU would be of order $ a \sim 2 \times 10^{-19}$ cm/s$^2$, corresponding to a radial shift of roughly 1 cm over a 10-year timescale. This small shift is undetectable with our proposed network.

\bibliography{}
\bibliographystyle{aasjournal}



\end{document}